\documentclass[12pt]{article}
\usepackage{amsmath}
\usepackage{amsfonts}
\usepackage{graphicx}
\usepackage{xcolor}
\usepackage{authblk}
\graphicspath{ {images/} }
\usepackage[margin=1in,top=1.5in,bottom=1.5in]{geometry}

\author[1, 2]{Sara A. Safari}
\author[1]{Christof Schmidhuber}
\affil[1]{Zurich University of Applied Sciences\\
School of Engineering, Technikumstrasse 9\\
 CH-8401 Winterthur, Switzerland}
\affil[2]{University of Zurich\\ Department of Mathematical Modeling and Machine Learning ($DM^3L$)\\ Winterthurerstrasse 190\\ CH-8057 Zurich, Switzerland}

\begin{document}

\title{Trends and Reversion in Financial Markets\\ [2pt]
 on Time Scales from Minutes to Decades}

\maketitle
 
\thispagestyle{empty}

\newpage
\begin{abstract}
We empirically analyze the reversion of financial market trends with time horizons ranging from minutes to decades. 
The analysis covers equities, interest rates, currencies and commodities and combines 14 years of futures tick data, 
30 years of daily futures prices, 330 years of monthly asset prices, and yearly financial data since medieval times. \\

Across asset classes, we find that markets are in a {\it trending regime} on time scales that range from a few hours to a few years, 
while they are in a {\it reversion regime} on shorter and longer time scales. In the {\it trending regime}, weak trends tend to persist, 
which can be explained by herding behavior of investors. However, in this regime trends tend to revert before they become strong enough to be statistically significant, 
which can be interpreted as a return of asset prices to their intrinsic value. 
In the {\it reversion regime}, we find the opposite pattern: weak trends tend to revert, while those trends that become statistically significant tend to persist. \\

Our results provide a set of empirical tests of theoretical models of financial markets. 
We interpret them in the light of a recently proposed lattice gas model, where the lattice represents the social network of traders, 
the gas molecules represent the shares of financial assets, and efficient markets correspond to the critical point. 
If this model is accurate, the lattice gas must be near this critical point on time scales from 1 hour to a few days, with a correlation time of a few years.

\end{abstract}

\thispagestyle{empty}
\newpage
\setcounter{page}{3}

\section{Introduction}
\label{section:intro}

Trendfollowing has been one of the most popular trading strategies in financial markets at least since the success of the turtle traders in the 1980's \cite{turtles}. 
Market trends have also been studied in the literature since the 1990's.
Early studies such as  \cite{cutler,silber} have shown that trend-following strategies can consistently generate risk-adjusted returns that exceed ordinary risk premia. Alternative beta strategies have tried to systematically replicate them \cite{fung, jaeger}, 
and many analyses, such as \cite{erb, miff, shen, mosk, menk, baz}, have demonstrated that these excess returns arise across all asset classes,
have low correlation with traditional asset returns, and complement other systematic strategies well. 
More recently, it has also been shown that market trends have already existed at least since the 19th century \cite{hurst, lemp, grey}. \\

Traditional trend-followers are ``long'' a given market when its current trend is positive, and ``short'' when the trend is negative. 
The implicit assumption is that the future expected return in a given market has the same sign as the current trend in that market.
This contrasts with the intuitive expectation that a trend should revert once it has taken the market price 
of an asset too far away from its intrinsic value. This trend reversion has been quantitatively observed in \cite{lemp} and 
has been modelled in \cite{black}. \\

As reported in \cite{schmidhuber}, we have independently confirmed and refined this observation by a precise empirical measurement of the future expected return in a market as a function of its current trend strength.
The key to filtering out the noise and getting high precision results was to aggregate not only across 30 years of daily futures returns and across a diversified set of 
24 equity indices, interest rates, currencies and commodities, 
but also across 10 trend horizons ranging from a few days to a year, as will be reviewed in more detail section 2. The latter aggregation makes sense, as it has long been known that similar trend-following strategies
tend to work over this wide range of time scales.\\

\begin{figure}[t]\centering
	\includegraphics[height=5.5cm]{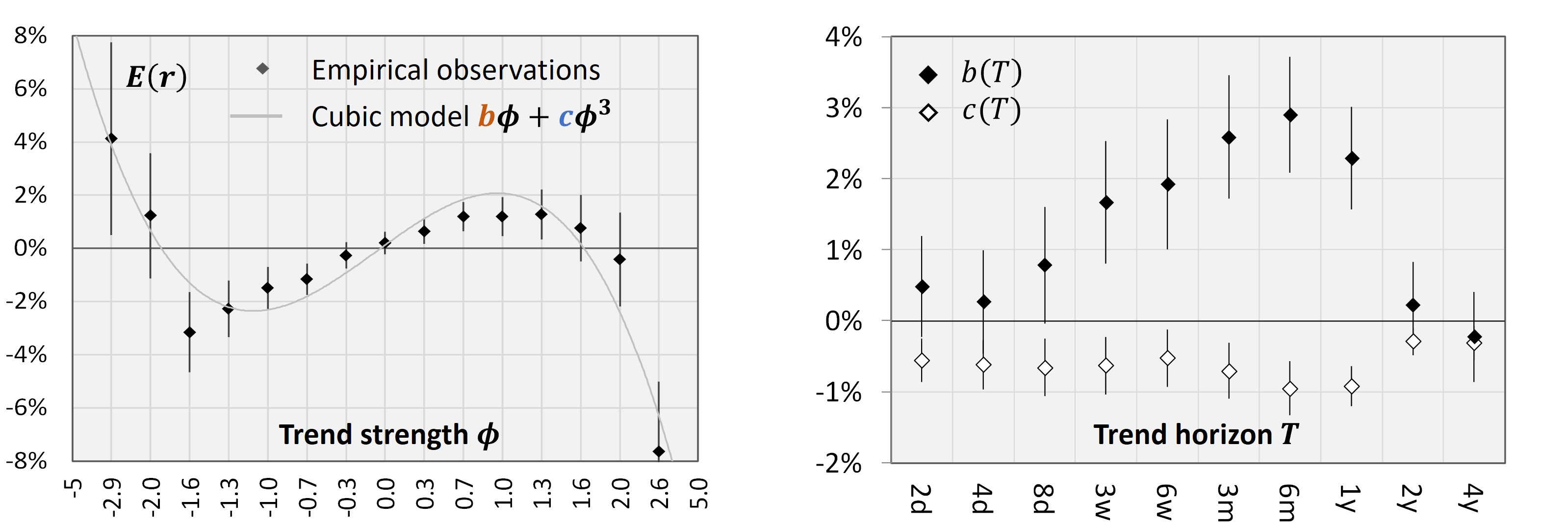}
	\caption{
{\it Left}: The expectation value $E(r)$ of the next day’s return of a futures market can be modeled by a cubic polynomial of the $t$-statistics $\phi$ of the trend. The linear term $b\phi$ models trend-persistence. 
The cubic term $c\phi^3$ with $c<0$ models trend-reversion.  
{\it Right}: The coefficients $b$ and $c$ depend on the time horizon $T$, which runs from 2 days to 4 years.
Both figures are aggregated across 24 assets and 10 time horizons as discussed in the text.}\label{figA}
\end{figure}

A key result of this previous study \cite{schmidhuber}, aggregated across assets and trend horizons, is shown in fig. 1 (left): 
trends tend to revert before they become statistically strongly significant. In other words, by the time a trend has become so obvious that everybody can see it in a price chart, it is already over. This is consistent with the hypothesis that any obvious market inefficiency is quickly eliminated by investors. More precisely, for a given market and a given time horizon, 
we have defined the strength $\phi$ of a trend as its $t$-statistics. 
We have found that tomorrow's return $R(t+1)$ (normalized to have variance 1) is modeled
well by a cubic polynomial of today's trend strength:
\begin{equation}
R(t+1)\ =\ a\ +{b}\cdot \phi(t)+{c} \cdot \phi(t)^3+\epsilon(t),\label{quartic}
\end{equation}
\noindent
where $a$ is a risk premium, $\epsilon$ represents random noise, and $b$ and $c$ are kinetic coefficients. A similar model had already been proposed in \cite{black} and extended in \cite{bouchTV}.
We have interpreted $b$ as the persistence of trends, and $c$, which is negative, as the strength of trend reversion, measured them quite precisely, 
and refined our observations as follows:
\begin{itemize}\addtolength{\itemsep}{0 pt} 
\item
The parameters $b, c$ are universal in the sense that they seem to be the same for all assets within the limits of statistical significance. 
This is in line with the fact that many successful trend-followers use the same systematic trading strategy for all assets. 
\item
$c$ is negative and does not seem to depend much on the trend's time horizon $T$. However, $b$ depends on $T$: 
it peaks at $T\sim$ 3-12 months, which is in line with the time scales on which trendfollowers typically operate. $b$ decays for longer or shorter horizons and
may become negative for $T<1$ day or $T>$ several years (fig. 1, right).
\item
While $c$ has been fairly stable over time, $b$ appears to have vanished over the decades.
This is in line with the fact that trendfollowing no longer works as well today as it did in the 1990's.
Markets seem to have become quite efficient with respect to trends.
\end{itemize}

Studying the interplay of trends and reversion in financial markets is not only of interest for traders.
It also holds clues for developing an accurate model of the markets that can replicate these and other empirically observed properties,
such as the so-called ``stylized facts'' of finance \cite{mant,cont}.
Such a comprehensive financial market model is in turn needed for better risk management, including monitoring and managing systemic risk. \\

Indeed, the above empirical observations have led us to propose a lattice gas model of financial markets \cite{me}.
The lattice represents the social network of investors, while the gas molecules represent the shares of an asset. 
The order parameter, i.e., the deviation of the molecule density from its mean, is identified with the deviation of the asset price from its intrinsic value.
Market efficiency implies a form of self-organized criticality \cite{bak}: one expects arbitrageurs to drive this lattice gas to its critical point, where a second-order phase transition occurs.
Near this point, the model can indeed explain the stochastic process (\ref{quartic}), where the cubic driving force is the derivative of the so-called Landau potential
known from the theory of phase transitions. This model can also explain other observed analogies between financial markets and critical phenomena,
including the universality of the coefficients $b,c$.\\

The quantitative predictions of such a network model of financial markets depend on the precise properties of the network, such as its topology or its fractal dimension.
Different network models predict different scaling behaviors of the markets under a rescaling of the time horizon $T$.
For example, if a simple (and unrealistic) hypercubic network topology was assumed, then a fractal dimension 
of $D\approx3$ could
quantitatively explain the observed value of the second Hurst coefficient of market returns \cite{me}.  
The model predictions for more realistic network topologies, such as in \cite{me2}, are being worked out separately. \\

In the current paper, we focus on establishing the empirical facts, i.e., the observed
 scaling bahaviour that such theoretical models must reproduce:
we extend the previous empirical measurement of the parameters $b, c$, as plotted in fig. 1 (right), in two directions:
\begin{itemize}
\item to intraday scales from 2 minutes to 1 day, using tick data from futures trading
\item to long-term scales from years to decades, using financial data since medieval times,
thereby extending previous historical results \cite{hurst,lemp,grey,black}.
\end{itemize}

This paper is organized as follows. In section 2, we briefly review the methodology already applied in \cite{schmidhuber} to daily data. In section 3, we extend these results to intraday data. 
In section 4, we extend them to long-term monthly data. Section 5 combines and summarizes our observations and interprets them within the framework of the lattice gas model of \cite{me}.

\section{Review: Analysis of Daily Data}
\label{section:review}

To empirically measure the interplay of trends and reversion in financial markets, we follow the methodology of \cite{schmidhuber} based on the data \cite{mendel}, which we briefly review in this section. 

\subsection{Data Set}
\label{section:2.1data}

The analysis of daily and intra-day market data is based on historical log-returns of a set of futures prices. 
We use futures prices, instead of the underlying market prices, because futures prices are guaranteed to be marked-to-market every day. 
Our set of futures is diversified across asset classes (equity indices, interest rates, currencies, commodities), regions (Americas, Europe, Asia) and commodity sectors (energy, metals, agriculture). \\

For daily closing prices, we have used the broadly diversified set of liquid contracts in table 1 over the period 1990-2020 \cite{mendel}.
For intraday data, we will use end-of-minute prices of a similar set of futures contracts, as described in section 3.
For long-term data covering several centuries, we will use end-of-month prices of a diversified set of assets, as described in section 4,
since the histories of futures prices are insufficient. In addition, for an indicative analysis of very long-term time horizons, we will use yearly data since medieval times for some assets in section 4.

\begin{table}[h!]\centering
\begin{tabular}{ |p{4.3cm}||p{3.3cm}|p{3.3cm}|p{3.3cm}|  }\hline
	{Asset classes vs. regions}& America  &Europe &Asia\\	\hline\hline
	Equities   & {S\&P 500}    &DAX 30&  { Nikkei 225}\\
	&  TSE 60  & { FTSE 100 }  & Hang Seng\\ \hline
	Interest rates &{US 10-year} & { Germ. 10-year} &  Japan 10-year \\
	&{Can. 10-year}  & {UK 10-year} &  Australia 3-year \\	\hline
	Currencies&   {CAD/USD}  &  { EUR/USD} & { JPY/USD}\\
	&   NZD/USD & {GBP/USD }  &AUD/USD\\	\hline\hline
	Commodities& { Crude Oil } &{ Gold} &{ Soybeans}\\
	& {Natural Gas}  & Copper &Live Cattle\\	\hline\hline
	Com.-Sectors:  & Energy  &Metals   &Agriculture\\	\hline
\end{tabular}
\caption{List of financial markets used in the analysis of daily data.}	
\end{table}

\subsection{Trend Definition}

For each market $i$ and each time interval, we consider the closing prices $P_i(t)$. We define normalized log-returns $R_i(t)$: 
\begin{equation}
R_i(t)={r_i(t)\over \sigma_i}\ ,\ \ \ r_i(t)=\ln {P_i(t)\over P_i(t-1)}\ ,\ \ \ \sigma_i^2=\text{var}(r_i)\ ,\ \ \ \mu_i=\text{mean}(r_i)\ ,\label{return}
\end{equation}
where the long-term risk premium $\mu_i$ (used below) and the long-term standard deviation $\sigma_i$ of a market $i$ are measured over the whole time period
(to avoid biases, in the out-of-sample cross-validation of our results, the risk premia and variances are estimated only from the training samples, excluding the validation samples).
For each market and each time interval, we compute different trend strengths on different (partly overlapping) time scales $T$:
\begin{itemize}\addtolength{\itemsep}{0 pt} 
\item Minute data: $T_k=2^k$ minutes with $k\in\{1,2,3,...,10\}$ ($2^6$ minutes $\approx1$ hour)
\item Daily data: $T_k=2^k$ business days with $k\in\{1,2,3,...,10\}$ ($2^8$ trading days $\approx1$ year)
\item Monthly data: $T_k=1.5\cdot 2^k$ months with $k\in\{1,2,...,8\}$ ($1.5\cdot2^3$ months $=1$ year)
\item Yearly data: $T_k=2^k$ years with $k\in\{1,2,3,...,7\}$ (based on $800$ years of data)
\end{itemize}

At a given point in time, a market may well be in an up-trend on one scale, but in a down-trend on another scale.  
For a given horizon $T$, we define the trend strength $\phi_{i,T}(t)$ of a given market $i$ at the close of trading of interval $t\in Z$ as a weighted average of previous log-returns of that market in excess of the long-term risk premium:
\begin{equation}
\phi_{i,T}(t)=\sum_{n=0}^\infty w_T(n)\cdot \hat R_i(t-n)\ \ \ \text{with}\ \ \ \hat R_i(t-n)=R_i(t-n)-{\mu_i\over\sigma_i},
\end{equation}
where $w_T(n)$ is a weight function. For monthly data, we use a variant of these trend strengths, where the variances are computed by an exponentially weighted moving average over the preceding $30$ years.  
We normalize the weight function $w_T(n)$ such that the trend strength $\phi_{i,T}$ has standard deviation 1. Assuming that returns over non-intersecting intervals are independent of each other (which is true to high accuracy), this implies: 
\begin{equation}
\sum_{n=0}^\infty w_T^2(n)=1.
\end{equation}
With this normalization, $\phi_{i,T}$ is the t-statistics of the trend. 
All trend strengths are then comparable with each other, so we can aggregate across all markets and time scales.
To reduce the effect of outliers, we put a ceiling/floor on the trend strength at $\pm 2.5$.
We also put a ceiling/floor on returns that are more than 20 standard deviations from the mean.\\

A popular weight function corresponds to a moving average cross-over: the trend strength is proportional to the average price over a short time period $S$
minus the average price over a long time period $T$, corresponding to the wedge in fig. 2 (left, solid line). 
For our study, we use the following weight function for the trend strength $\phi_T$, which is similar but smoother: 
\begin{eqnarray}
\tilde w_T(n)&=&M_T\cdot\ n\cdot e^{-2n/T} \ \ \ \text{with normalization factor} \ \ M_T.
\end{eqnarray}
The "average lookback period" of $\phi_T$, i.e., the expectation value $E[n+1]$ of the number of intervals we look back (where "today" is $n=0$), is then $T$.
While the precise definition of the weight function does not matter much, we use this one, because it
has only one single free parameter $T$ (which reduces the risk of overfitting historical data), and because it can efficiently be computed recursively.\\

Table 2 shows a small excerpt of the resulting data table. 
It contains $N\times24\times10$ pairs of trend stregths on a given day, where $N$ is the number of days, 24 is the number of assets, and 10 is the number of trend horizons.
The response variable is the next-day return. 

\begin{table}[h!]\centering
\begin{tabular}{ |p{1cm}|p{2.5cm}||p{1.5cm}|p{1.5cm}|p{1.5cm}|p{1.5cm}|p{2.5cm}|  }	\hline
 	\multicolumn{2}{|l|}{Time / Market} & \multicolumn{4}{|l|}{Trend strength at given day}&Return\\ \hline
	{Day}& Asset & k=1 & k=2& ... & k=10& (next day) \\	\hline
	1 & FTSE 100& -1.2&-0.8 &... &0.7 &1.7\\  
	1 & EUR/USD & 0.2& 0.4 & ... & 0.7&0.1\\	
	1 & Crude oil & 0.9 & 1.4 &... &0.8 &-2.1\\ \hline
	2 &  FTSE 100& -0.6& -0.6&...  & 1.1 &-0.4 \\  
	2 & EUR/US & 0.5& 0.5& ... & 1.1&0.8\\	
	2 & Crude oil & 1.4&1.6 & ... &1.1 &1.4\\ \hline
	3 & ... &  ... &  ... &  ... &  ... &  ... \\ \hline     
\end{tabular}
\caption{Small excerpt of the data table used in the daily data analysis, showing two trading days (out of 30 years of data), and 3 of the 24 assets.
For each asset and each day, 10 different trend strengths are computed, measuring the trend over the past $T=2^k$ days with $k\in \{1,2,...,10\}$. The last column 
is the response variable, the asset return on the day after.}	
\end{table}

\subsection{First Look at the Data}

For a first explorative data analysis, we define 15 buckets $b_k$ of similar trend strength 
\begin{equation}
\phi\in b_k\ \ \ \text{with}\ \ \ k\in\{-7,+7\},\ \ \ b_k=[\ {k\over3}-{1\over6}\ ,\ {k\over3}+{1\over6}\ [\ .\label{buckets}
\end{equation}
I.e., bucket $b_0$ covers trend strengths from $-1/6$ to $+1/6$, $b_1$ from $+1/6$ to $+1/2$, and so on.
We extend the outermost buckets $b_{\pm7}$ to $\pm\infty$.
We then measure the average future return $r(t+1)$, if $\phi(t)$ lies in a given bucket on day $t$.
For daily data, fig. 1 (left) shows this average next-day return for a bucket as a function of the average trend strength in that bucket.
The results are aggregated across 30 years of daily returns for each market, and across all 24 markets.
Unlike previous studies, \cite{schmidhuber} also aggregates across different time scales $T$. 
This was key to clearly observing the statistically highly significant patterns in fig. 1. 
For shorter time windows, or for any single time scale, these patterns are too blurred to be detected reliably. 
This may explain why our findings  have not been reported already a decade ago. \\

This first visualization of the data tells us what kind of pattern to look for:
it suggests to model tomorrow's market return as a polynomial of today's trend strength,
which should contain at least the linear and the cubic terms. 

\subsection{Regression Analysis}

To prove and quantify this suspected pattern, we dissolve our trend strength buckets (\ref{buckets}) again
and perform a polynomial regression on the full underlying data set of
$24\cdot260\cdot30\cdot10\approx2$ million data pairs (24 markets, 260 trading days per year, 30 years, 10 trend strengths).
The following results are thus independent of the definition of buckets in (\ref{buckets}).
We regress the next-day normalized log-return $R_i(t+1)$ (\ref{return}) 
against a polynomial of the current trend strength $\phi_{i,k}(t)$ across all markets $i$ and time scales $T=2^k$.
\begin{equation}
R_i(t+1)=a+b\cdot  \phi_{i,k}(t)+c \cdot \phi_{i,k}^3(t)+\epsilon_i(t+1),\label{cubic}
\end{equation}
where $\epsilon$ represents random noise, and $a$ measures the average normalized risk premium $\mu_i/\sigma_i$ across all assets. 
We ignore $a$ and focus on determining the 
persistence of trends $b$, and the strength of trend reversion $c$. 
The results of the overall regression analysis are shown in table 3. 
As reported in the appendix, the $\phi^2$ term is not statistically significant, and adding higher-order terms in $\phi$ would also reduce
the statistical significance of the model.  \\

\begin{table}[h!]\centering
\begin{tabular}{ |p{2.5cm}||p{3.5cm}|p{2.5cm}|p{2.5cm}|  }	\hline
 	Parameter & \multicolumn{3}{|l|}{Nonlinear regression results} \\ \hline
	{Coefficient}& Value  &Error &t-statistics\\	\hline
	$a$   & $1.33$\%    &$\pm 0.41$\% &  $3.3$\\  
	$b$ &$1.29$\% & $\pm 0.43$\% &  $3.0$ \\	
	$c$ &  $-0.62$\%  &  $\pm 0.23$\% & $2.7$\\ \hline\hline
	{R-squared} & Single time scales & \multicolumn{2}{|l|}{Aggregated across time scales} \\ \hline
	$R^2$ &$1.31$\ bp & \multicolumn{2}{|l|}{$4.91$\ bp}  \\	
	$R^2_{adj}$ &  $1.03$\ bp  &  \multicolumn{2}{|l|}{$3.98$\ bp}  \\ \hline
\end{tabular}
\caption{Regression results for daily returns: $a$ is the intercept, while $b$ and $c$ are the regression coefficients of the two explanatory variables, the trend strength and its cube.}	
\end{table}

The coefficients $a,b,c$ and the R-squared's are very small, as they {\it must} be: 
if the patterns were stronger, traders could achieve stellar Sharpe ratios by exploiting them. In fact, it is shown in \cite{schmidhuber} that 
the measured coefficients are just big enough to explain the Sharpe ratios of order 1 that trend followers have historically achieved.
$b$ and $c$ are statistically highly significant out-of-sample, although they seem to have decreased over the decades.
Within the limits of statistical significance, we find that $b$ and $c$ are universal, i.e., the same for all assets. 
While the strength of reversion $c$ seems to be approximately constant, we find that the persistence of trends $b$ depends on the time horizon of the trend. 
Within the range of time scales considered here, and averaged over the past 30 years, we have found 
that trends only persist if the time horizon ranges from a few days to one or two years. 

\subsection{Statistical Analysis}

\begin{figure}[t]\centering
	\includegraphics[height=5.4cm]{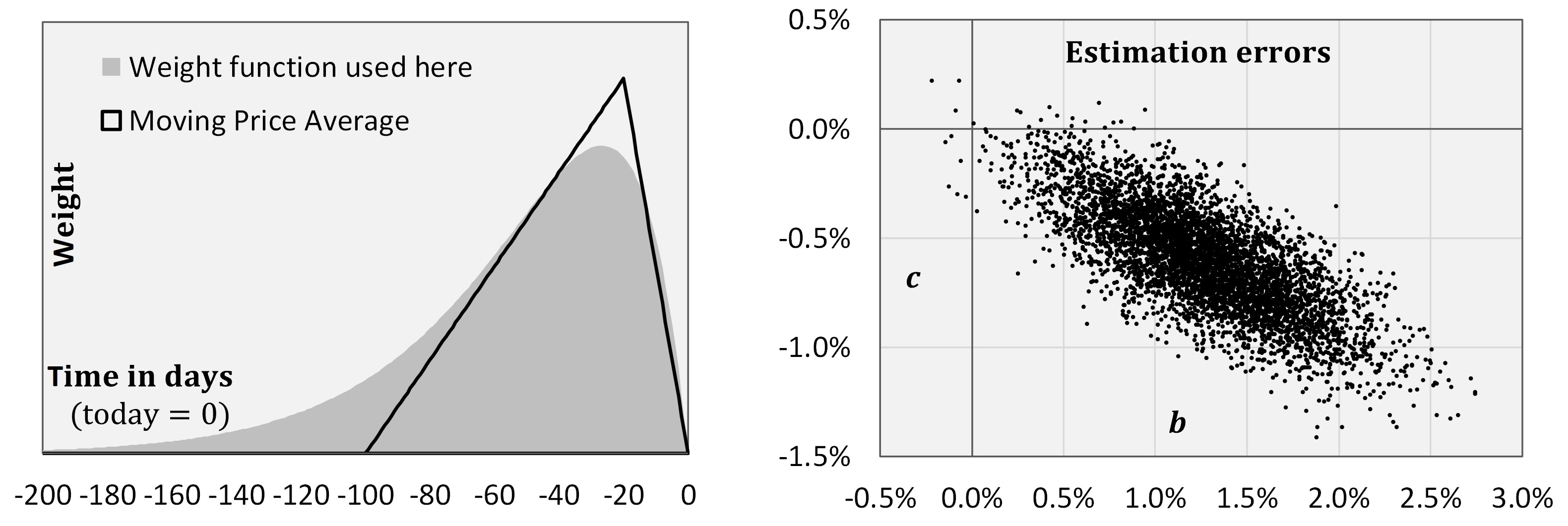} 
	\caption{Left: Our trend strength is a weighted sum of past log-returns, similar to a moving average crossover.
Right: regression coefficients $b,c,$ for 5000 bootstrapping samples.}
	\label{figA}
\end{figure}

Given the small values of the coefficients, and the huge amount of noise in financial market data, we must assess the statistical significance of the results very carefully. 
Since market returns cannot be assumed to be independent, identically distributed normal random variables, we cannot trust standard formulas for the t-statistics, adjusted R-squared's, and other statistics. Instead, the test statistics in table 2 have been measured empirically as follows:
\begin{itemize}
\item  The error of the coefficients and their t-statistics were computed by bootstrapping, using 5000 bootstrapping samples. Regression on each new sample of days yielded the distribution of the 5000 regression coefficients $b$ and $c$ shown in fig. 2 (right), from
which the estimation errors in table 3 were read off. These actual errors are 3-4 times as large as the errors that standard regression tools would report. Nevertheless, even with these large errors the regression coefficients are still statistically highly significant. 
\item The adjusted R-squared was computed by 15-fold cross validation. The resulting out-of-sample R-squared is reported as $R^2_{adj}$ in table 3. 
The actual out-of-sample correction $R^2-R^2_{adj}$ is about 3 times as big as what standard regression tools would report. Nevertheless, the out-of-sample $R^2_{adj}$ is still highly significant despite being very small, and 
agrees with the Sharpe ratio that has been achieved by trend-followers in practice. 
\item  
Table 3 also reports ${R}^2$ and ${R}^2_{adj}$ "aggregated across time scales". Those are based on using the equally-weighted mean of the 10 trend strengths on each day to predict the next-day return for each market. I.e., we combine the 10 different trend factors into a single one, which has a higher predictive power than each single factor by itself. 
\end{itemize}

The regression results of table 3 and further regression analyses in \cite{schmidhuber} confirm and quantify our conclusions from fig. 1. In particular, we see that the values of $b$ and $c$ - although very small - are statistically highly significant, despite the fact that market returns are neither normally, nor identically, nor independently distributed. \\

The reported values of $b, c$ are averaged over the past 30 years. In more recent time periods, our measurements indicate that at least $b$ has decreased. 
E.g., section 4.4 of \cite{schmidhuber} confirms this at the 97.5\% confidence level, based on a linear decay ansatz, and estimates that $b\rightarrow0$ somewhere between 2010 and 2030.  
This has a natural interpretation: so much money has flown into trend-following
that it does not work as well any more as it used to. 

\section{Analysis of Intraday Data}

We now extend the previous analysis from daily data to intraday data with time scales ranging from 2 minutes to 1 trading day, using tick data from world-wide futures markets
(i.e., data that record every buy order, every sell order, and every trade, including a precise timestamp).
For previous work on intraday trends see, e.g., \cite{olson,daco}.

\subsection{Data Set}

As in \cite{mendel}, our analysis is based on historical log-returns of futures contracts that are rolled 5 days prior to first notice
(i.e., the expiring contract is sold and a new contract is bought). 
We again diversify across four different asset classes (equity indices, interest rates, currencies, and commodities) 
and three regions (Americas, Europe, and Asia) as shown in table 4. 
Compared with the analysis of daily data, we drop some markets that are less liquid or for which we have less clean data (Nikkei and Hang Seng indices, Japanese and Australian government bonds, AUD and NZD currencies, copper and live cattle). We also replace the DAX by the Eurostoxx 50 Index.
Our data were provided by TickData (https://www.tickdata.com) and cover the 14-year period 2010-2023. 

\begin{table}[h!]\centering
\begin{tabular}{ |p{3.3cm}||p{2.5cm}|p{2.5cm}|p{2.5cm}|p{2.5cm}|  }	\hline
	{Asset Class}&{\it Market 1}   &{\it Market 2} &{\it Market 3}&{\it Market 4}\\	\hline\hline
	Equities   & S\&P 500    & Eurostoxx 50& FTSE 100&  Nikkei 225\\ \hline
	Interest rates &US 10-year & Can. 10-year & Ger. 10-year   & UK 10-year \\	\hline
	Currencies&   CAD/USD  &  EUR/USD  & GBP/USD& JPY/USD \\	\hline
	Commodities & Crude Oil & Natural Gas   &Gold &Soybeans\\	\hline
\end{tabular}
\caption{List of financial markets used in the analysis of intraday (end-of-minute) data.}	
\end{table}

The analysis is based on end-of-minute prices (last traded price in each minute), rather than the whole set of intra-minute tick data.
These end-of-minute prices are sufficient for our purposes, as we do not analyze intra-minute time scales here.   
From the prices, we compute the returns following the definition (\ref{return}), normalizing them in the same fashion as in \cite{mendel} to have standard deviation 1. \\

The analysis of intraday data must account for the overnight jump in market prices due to the closure of the exchange and re-opening on the next trading day. 
As a result, the return in the opening minute is typically quite large, which can distort our analysis. 
Moreover, in the absence of overnight data, no meaningful trend strength can be computed during the night and in the early trading day.
We therefore analyze the intraday trending behavior for trend horizons of $2^k$ minutes with $k\in\{1,2,...,6\}$ ($k=6$ corresponding to roughly 1 hour) on a day-by-day scheme: 
the time series of returns are broken up into sub-sets, one for each day, from the opening to the close of each market in its respective time zone. On each day, the first-minute return is ignored, and the first $T$ minutes of the day serve to build up the $T$-minute trend strength.
After this ramp-up period, the trend strengths of the different horizons at the end of each minute as well as the next-minute return are measured. \\

We then aggregate the resulting tables of trend-strengths and returns (6 trend strengths and 1 next-minute return for each minute) over all days to increase the statistical significance of our results.
We also aggregate across the different markets, which are traded in different time zones. Hereby, we implicitly assume that the intraday trending behavior of markets is independent of the 
asset class and the time zone, an assumption that we subsequently test. Based on these data tables, we perform the regression. 
As a further cross-check, we subsequently test whether our regression results change significantly if we insert a time lag, i.e., if we replace the next-minute return by the next-next-minute return.\\

This day-by-day analysis limits the trend horizons $T=2^k$ minutes to at most a few hours. 
For $k\ge7$, we instead use continuous time series of minute returns through the whole 14-year period, reinstating
the first-minute returns to include the overnight returns. When connecting minute results with daily results, we interpret $2^{10}$ minutes as one trading day. 
This accounts for the fact that there is no or little overnight trading in many markets.

\subsection{First Look at the Data}

As for daily returns, we first visualize the data to decide how to set up the regression. 
We first focus on trend strengths in the range $[-2.5,+2.5]$ and on the 3 short time horizons 2, 4, and 8 minutes.
Fig. 3 (left) shows the scatter plots of the average next-minute return as a function of the trend strength at the end of the minute before, based on trend strength buckets as for daily data.
The results are aggregated over all years and across all assets.\\

\begin{figure}[t]\centering
    \includegraphics[height=5.7cm]{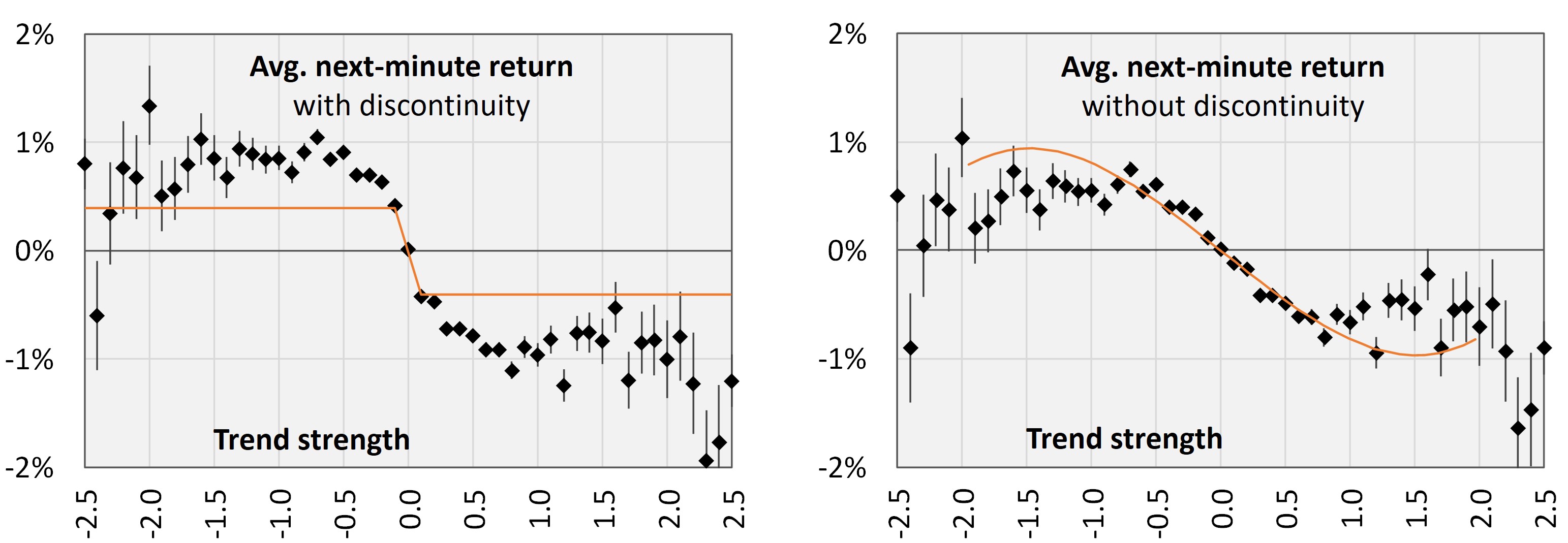}
	\caption{Average next-minute return as a function of the current trend strength for horizon $T=2^k$ minutes with $k=1,2,3$. 
	Left: the discontinuity at the origin due to the tick size is modelled by a step function. Right: the remaining pattern is modelled by a polynomial.}
	\label{fig:minute_w}
\end{figure}

The average next-minute return is mostly positive for negative trend strength and negative for positive trend strength, so futures prices tend to revert at minute-scales.
There is a discontinuity at trend strength 0. We have found that it does not go away if we make the trend strength buckets even smaller, and 
that it decreases with the trend horizon $T$ like $1/\sqrt T$.
This discontinuity has a natural interpretation: futures prices are not continuous but discrete with a certain tick size.
E.g., S\&P 500 futures are priced in units of 25 cent (e.g., at \$5012.00 or \$5012.25, but not in between).
When the (compounded) underlying market price is in the middle between two ticks, the futures price tends to oscillate between them, leading to a 
negative auto-correlation of market returns from one minute to the next. \\

The size of the discontinuity depends on the tick size and the liquidity of the market.
In the trend strength for horizon $T$, the weight of the previous minute is of order $1/\sqrt T$, which explains the decrease of the discontinuity with $T$.  
This effect can be used to reduce trading costs, but it can hardly be used for a profitable trading strategy. 
One could of course try to earn a so-called "market maker fee" by placing buy orders at the lower price just after an up-tick,
and placing sell orders at the higher price just after a down-tick. However, there are trading costs. Moreover, 
there is the risk that the price does not revert but runs away due to an underlying trend or a buy/sell order by a large investor.
We assume that markets are efficient in the sense that any trading profit just compensates for this risk.\\
 
\begin{figure}[t]\centering
    \includegraphics[height=5.5cm]{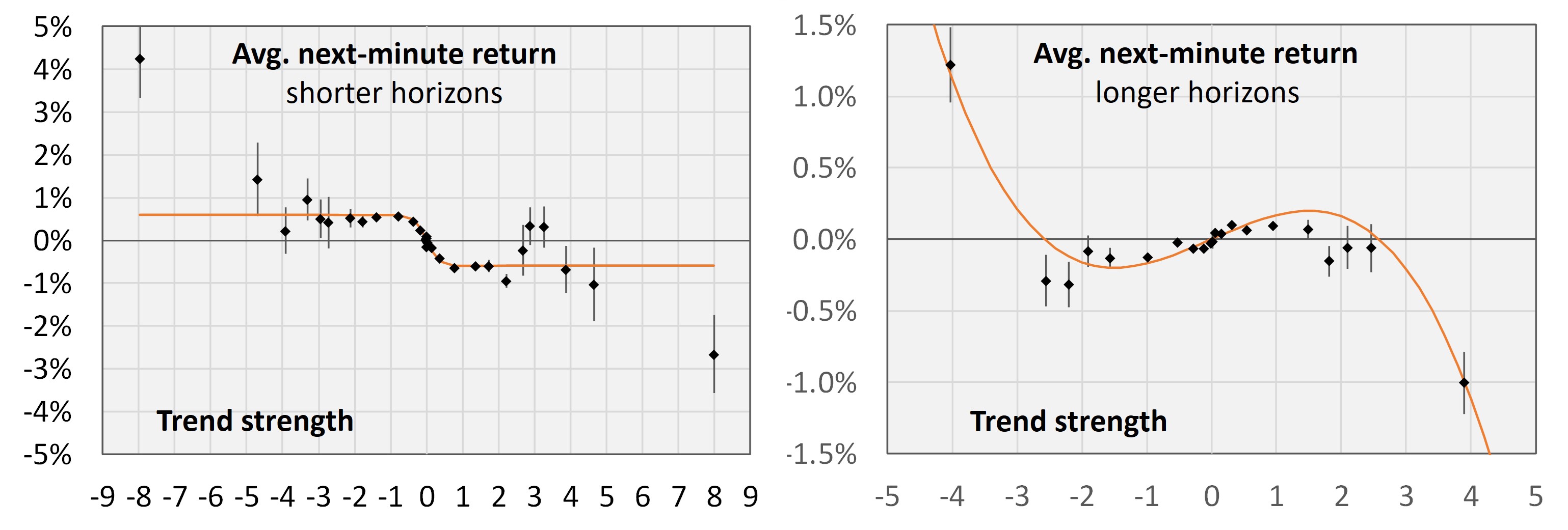}
	\caption{Left: extension of fig. 3 (left) to larger and smaller trend strengths after removing the discontinuity. 
	A hyperbolic tangent models the tails poorly for horizons $T\le4$. Right: analogous graph for $T=1$ hour to 1 day. The pattern is modelled well by a cubic polynomial.}
	\label{fig:minute_w}
\end{figure}

As seen from fig. 3 (right), after the discontinuity is removed, the remaining pattern could be modelled by a polynomial or a hyperbolic tangent, which can be taylor-expanded:
\begin{equation}
A\cdot\tanh Bt=AB\cdot t-{1\over3}AB^3\cdot t^3+{2\over15}AB^5\cdot t^5+...\label{taylor}
\end{equation}
For $\vert t\vert<2$, the linear and cubic terms are sufficient. However, their signs are inverted compared with the results from daily data:
Even without the discontinuity due to the tick size, we see that small trends tend to mean-revert (negative linear coefficient), while the reversion becomes weaker for stronger trends (positive cubic coefficient). \\

A possible explanation is that small random fluctuations of the prices of the underlying are quickly corrected by statistical arbitrage traders (i.e., traders who systematically sell/buy assets 
whose price has risen/fallen significantly, thereby driving the asset price back down/up).
However, in cases where there is an underlying cause for the price change,
such as an earnings announcement or an economic/political event, one expects that the trend becomes strong and persists until a new equilibrium price is reached that prices in this event.
Such situations should weaken the general pattern of mean reversion at minute scales.
Again, it would be difficult to profitably exploit these weak patterns in the face of trading costs, 
beyond earning a fair risk premium. In this sense, the patterns are still consistent with assuming that markets are efficient \cite{fama} at intraday scales.\\

Fig. 4 (left) differs from fig. 3 (right) by a more fine-grained resolution of the tails (with the discontinuity removed) for time horizons up to 16 minutes. 
It is clear that a (negative) hyperbolic tangent of the trend strength $t$ does not model the tails well, 
and that we must include at least the first three terms in (\ref{taylor}), i.e., we must fit the coefficients of $t,t^3,t^5$ idependently from each other.
On the other hand, for time horizons $T\ge 1$ hour, $t$ and $t^3$ seem to suffice to model the tails (fig. 4, right). 
In the following, we will therefore model the pattern by a cubic polynomial in $t$ for horizons $T=2^k$ with $k\ge6$,
while we add a quintic term for $k<6$. We will see that the quintic coefficient becomes insignificant for $k\ge5$.

\subsection{Regression Analysis}

To quantify these qualitative observations, we thus perform a regression with four factors:
linear, cubic and quintic terms $\phi,\ \phi^3$, and $\phi^5$, and the step function sign($\phi$): 
\begin{equation}
R(t+1)=a+b\cdot  \phi(t)+c \cdot \phi^3(t)+d \cdot \phi^5(t)+e \cdot \text{sign}\ \phi(t)+\epsilon(t+1),\label{quintic}
\end{equation}
The latter measures the size of the discontinuity and removes it.
The nonlinear regression results, aggregated across all assets and time scales $T=2^k$ with $k\in\{1,2,3,4\}$ are shown in table 5. 
As for daily data \cite{mendel}, the estimation errors were computed empirically based on 
bootstrapping samples, each of which contained a random selection of whole trading days (sampling the same days for all assets, with replacement).
The adjusted R-squared, reported here in basis points (0.01\%), has been estimated out-of-sample by  12-fold cross-validation. 

\begin{table}[h!]\centering
\begin{tabular}{ |p{2.5cm}||p{3.5cm}|p{3.5cm}|p{3.5cm}|  }	\hline
 	Parameter & \multicolumn{3}{|l|}{Regression results for trend horizons up to 16 minutes} \\ \hline
	{Coefficient}& Value  &Error &t-statistics\\	\hline
	$a$   & +0.017\%    &$\pm 0.020$\% &  +0.8\\  
	$b$ &$-0.912$\% & $\pm 0.067$\% &  $-13.6$ \\	
          $c$ &${+0.259\%}$ & $\pm 0.056$\% &  $+4.6$ \\
          $d$ &${-0.038\%}$ & $\pm 0.009$\% &  $-4.2$ \\	
          $e$ &${-0.282\%}$ & $\pm 0.023$\% &  $-12.5$ \\	 \hline\hline
	{R-squared} & Single time scales & \multicolumn{2}{|l|}{Aggregated across time scales} \\ \hline
	$R^2$ &$0.43$\ bp & \multicolumn{2}{|l|}{0.56\ bp}  \\	
	$R^2_{adj}$ &  $0.42$\ bp  &  \multicolumn{2}{|l|}{0.55\ bp}  \\ \hline
\end{tabular}
\caption{Regression results with minute data according to ansatz (\ref{quintic}) for short-term trends.}
\end{table}
 
\begin{table}[h!]\centering
\begin{tabular}{ |p{2.5cm}||p{3.5cm}|p{3.5cm}|p{3.5cm}|  }	\hline
 	Parameter & \multicolumn{3}{|l|}{Regression results for trend horizons $\ge1$ hour.} \\ \hline
	{Coefficient}& Value  &Error &t-statistics\\	\hline
	$a$   & +0.003\%    &$\pm 0.024$\% &  +0.1\\  
	$b$ &$+0.132$\% & $\pm 0.046$\% &  $+2.9$ \\	
          $c$ &${-0.039\%}$ & $\pm 0.014$\% &  $-2.8$ \\
          $e$ &${-0.071\%}$ & $\pm 0.018$\% &  $-3.9$ \\	 \hline\hline
	{R-squared} & Single time scales & \multicolumn{2}{|l|}{Aggregated across time scales} \\ \hline
	$R^2$ &$0.012$\ bp & \multicolumn{2}{|l|}{0.025\ bp}  \\	
	$R^2_{adj}$ &  $0.009$\ bp  &  \multicolumn{2}{|l|}{0.020\ bp}  \\ \hline
\end{tabular}
\caption{Regression results with minute data according to ansatz (\ref{quintic}) for longer-term trends.}	
\end{table}

We see from table 5 that the discontinuity $e$ at the origin due to the nonzero tick size is statistically highly significant.
The linear coefficient $b$ is also highly significant, but now it is negative. This quantifies 
our qualitative observation from the scatter plot and confirms that, overall, markets tend to mean-revert at intraday scales
even after discounting for the reversion due to the tick size. 
The cubic coefficient $c$ is also highly significant and positive, while it was negative in the case of daily data.
The quintic coefficient is negative and highly significant as well. Even terms were not significant enough to be included in the analysis, as explained in the appendix.\\

Next, we perform the regression aggregated across the longer time horizons $T=2^k$ minutes with $k\in\{6,7,8,9,10\}$,
this time without the quintic term as discussed. The results are shown in table 6.
We observe that the discontinuity has become smaller, as expected. The linear/cubic coefficients have switched signs
and are now in line with the pattern from daily data: weak trends tend to persist ($b>0$), while strong trends tend to revert ($c<0$).

\subsection{Refined Analysis by by Time Horizon}

Fig. 5 shows the regression results for the coefficients $b_T, c_T, d_T,$ and $e_T$ seperately for each time horizon of $T=2^k$ minutes with $k\in\{1,...,10\}$. 
The regression equation is now
\begin{equation}
R(t+1)=a_T+b_T\cdot  \phi_T(t)+c_T \cdot \phi_T^3(t)+d_T \cdot \phi_T^5(t)+e_T \cdot \text{sign}\ \phi_T(t)+\epsilon_T(t+1),\label{cubicT}
\end{equation}
with the quintic term only present for $k\le6$. For better visualisation for $k>4$, we define the modified coefficients
\begin{equation}
(\tilde a_T, \tilde b_T, \tilde c_T, \tilde d_T, \tilde e_T)\ =\ (a_T, b_T, c_T, d_T, e_T)\cdot\sqrt{T\over60}\label{redef}
\end{equation}
and plot $\tilde b_T,\tilde c_T$ instead of $b_T,c_T$.
These modified coefficients have a simple interpretation. In first approximation, the price performs a Gaussian random walk.
Multiplying both sides of equation (\ref{cubicT}) with $\sqrt{T/60}$ replaces the coefficients $(a,b,c,d)$ by $(\tilde a,\tilde b,\tilde c,\tilde d)$,  
and the 1-minute return $R$ by the $T$-second return $R_T$ (with $R_{60}=R$), whose variance is normalized to 1:
\begin{equation}
R_T(t+{T\over60})\equiv\ln P(t+{T\over60})-\ln P(t)\ \ \Rightarrow\ \  \langle R_T^2\rangle   ={T\over60} \langle R^2\rangle=1.
\end{equation}
So we use the $T$-minute trend strength to predict the returns over a $T$-second (instead of 1-minute) time period.
$\tilde b _T$ and $\tilde c_T$ describe how financial markets behave if all time scales are rescaled by the same factor.
If markets were completely scale invariant, $\tilde b _T$ and $\tilde c_T$ would be independent of $T$ within their estimation errors.\\

\begin{figure}[t]\centering
	\includegraphics[height=5.5cm]{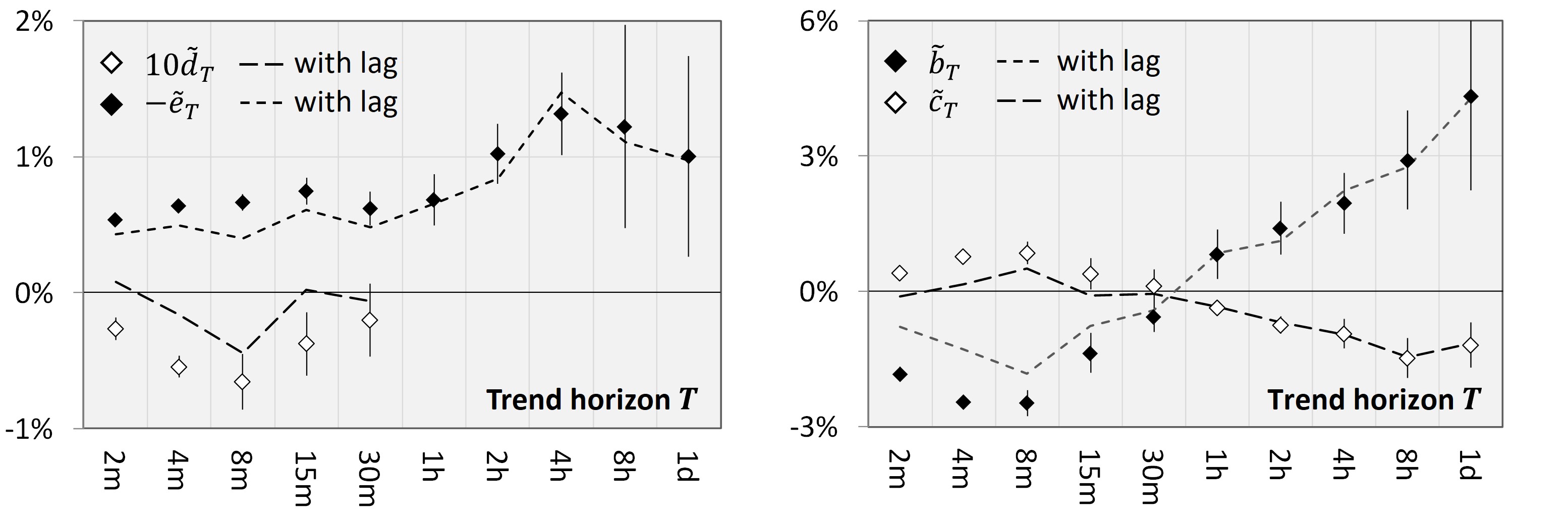} 
	\caption{Regression coefficients in ansatz (\ref{cubicT}) for trend horizons $T=2^k$ minutes, rescaled by $\sqrt T$. 
	Left: discontinuity $\tilde e_T$ due to tick size and quintic coefficient $\tilde d_T$ for short time horizons (rescaled by factors). Right: remaining linear and cubic coefficients $\tilde b_T$ and $\tilde c_T$.
     Dashed lines: same coefficients with a 1-minute time lag between measured trend and predicted return.}
	\label{fig:minute_w}
\end{figure}

We observe in fig. 5 that the rescaled tick size discontinuity is indeed negative and almost independent of $T$,
although it becomes statistically less significant for larger $k$ (fig. 5, left).
The quintic term is statistically significant and negative for time scales less than one hour.
Fig. 5 (right) shows that the regime where markets revert more strongly than implied by the tick size effect (negative $b$, positive $c$) ends at time scales of approximately 30 minutes.
For longer time horizons, weak trends tend to persist (positive $b$, negative $c$). $b$ becomes larger than $\vert e\vert$ at scales of a few hours,
indicating that these trends can even overcome the tick size reversion.
This "trending" regime extends up to daily scales, where our results connect with 
our previous results based on daily data \cite{mendel}, as seen in fig. 10 of section 5.\\

To verify that our results are not an artifact of any hidden auto-correlation of the returns due to unclean seperation of the trading minutes in our data,
the dashed lines in fig. 5 show $\tilde b_T, \tilde c_T$ with a 1-minute time lag inserted between the minute the trend is measured and the minute for which the return is predicted. 
Reassuringly, the results hardly change for $T\gg$ 1 minute. Of course, the 1-minute lag reduces the auto-correlations for shorter $T$. 

\subsection{Refined Analysis by Asset Class and Time Zone}

\begin{figure}[t]\centering
	\includegraphics[height=9.5cm]{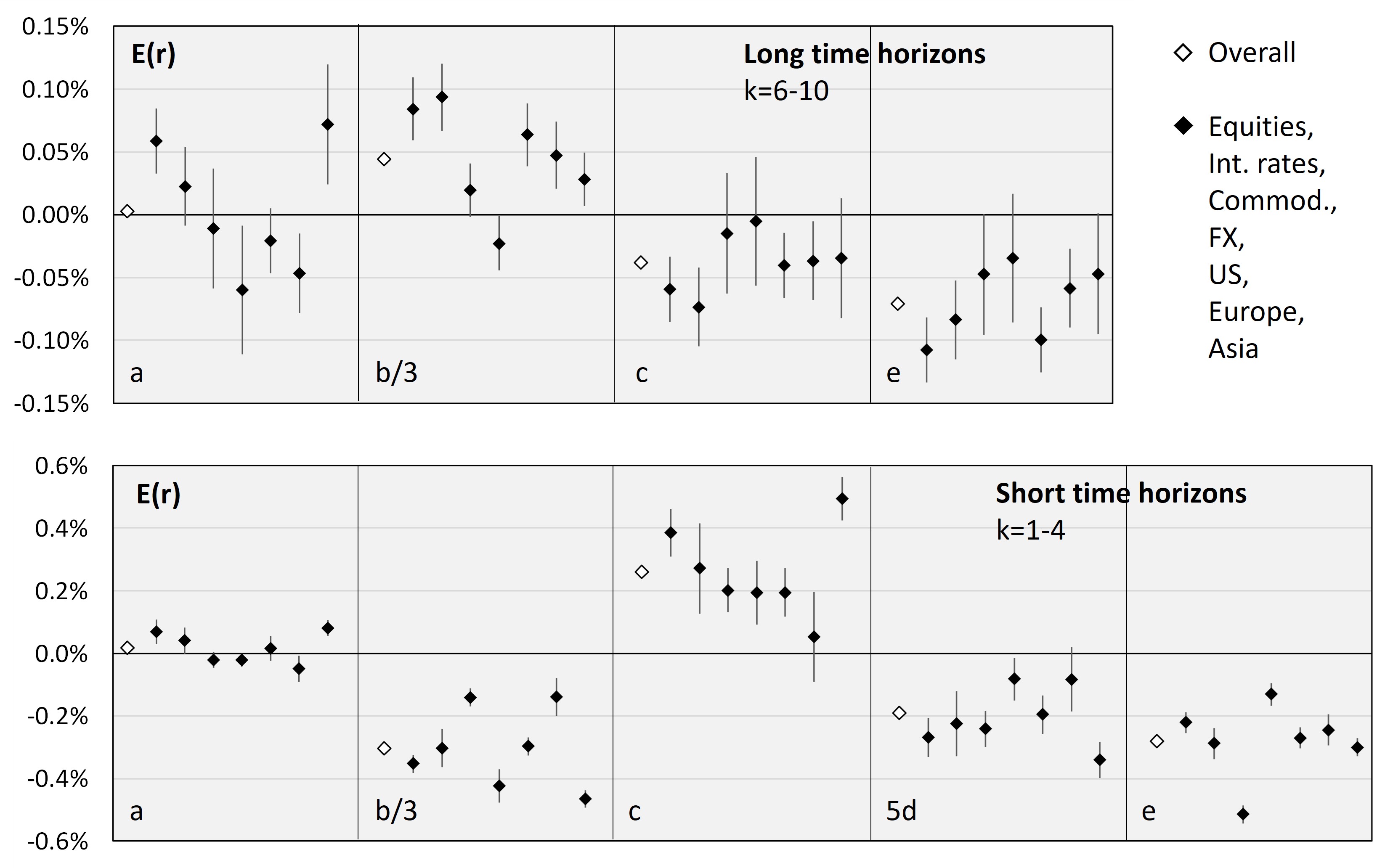} 
	\caption{Regression coefficients overall (white marks) and separately for equities, interest rates, commodities, currencies, and US/European/Asian trading hours (black).
          For time scales $\ge1$ hour (top), the values may be universal, but not for shorter time scales (bottom). }
	\label{fig:minute_w}
\end{figure}

So far, we have aggregated across different asset classes and time zones. We now perform the regression analysis separately for
\begin{enumerate}\addtolength{\itemsep}{-8 pt}
\item[(i)] Equity indices (all trading hours)
\item[(ii)] Interest rates (all trading hours)
\item[(iii)] Commodities (all trading hours)
\item[(iv)] Currencies (all trading hours)
\item[(v)] US markets (main trading hours 2-11 pm, all asset classes)
\item[(vi)] European markets (main trading hours 8am-2pm, all asset classes)
\item[(vii)] Asian markets (main trading hours 0-8 am, all asset classes)
\end{enumerate}

Fig. 6 shows the results for longer time scales $k=6,...,10$ (top) and for shorter time scales $k=1,2,3,4$ (bottom).
The white marks represent the overall values of the regression parameters, followed by their values for (i)-(vii) in this order (black marks).  
Some regression parameters have been rescaled to fit all of them in the same range. 
We observe that for horizons of one hour or more (longer time scales), the different values of each parameter more or less agree within the estimation errors.
In this sense, the parameter values may be universal. 
On the other hand, for time scales up to 16 minutes, the parameter values significantly differ from each other, 
depending on the asset class or time zone. Thus, at very short time scales, the parameter values are not universal. We return to this in scetion 5.

\section{Analysis of Centuries of Data}

While the previous section was dedicated to intraday time scales, this section measures the interplay of trends and reversion on very long time scales from a few years up to decades. 
This is based on monthly and yearly market prices. Here, we need not worry about time zones. Instead, the challenge is to collect and clean century-old financial market data.  \\

Trends in past centuries have previously been analyzed. 
In \cite{hurst}, it was found that trend-following on time scales from 1 month to 1 year 
would have been highly profitable across all asset classes since the 1880. 
In \cite{lemp}, this result was extended back to 1800. According to \cite{grey}, trends may even have existed for at least 800 years.  
After the analysis in this section was completed, we noticed that \cite{black} extends the 200-year analysis of \cite{lemp} to trend horizons of up to 5 years,
 including the cubic and also a quadratic term in ansatz (\ref{quartic}). 
The independently obtained results that we report below are consistent with those of \cite{black}. They extend them
 to more than 3 centuries, and to trend horizons of several decades. 

\subsection{Data Set}

We use monthly data from the “Global Financial Database”, covering the diversified set of markets shown in table 7. 
The time series we use all end in 2023, but they have different starting years, shown in brackets. 
We treat gold as a currency rather than a commodity, measuring the values of all currencies in units of the gold price. \\

\begin{table}[h!]\centering
\begin{tabular}{ |p{3cm}||p{1.6cm}|p{1.6cm}|p{1.6cm}|p{1.6cm}| p{1.6cm}|p{1.6cm}|}	\hline
	Asset Class&{\it 1}   &{\it  2} &{\it  3}&{\it  4}&{\it  5}&{\it 6}\\	\hline\hline
	Equity  & US  & UK & Germany &  France& Canada & Japan\\ 
	indices  & (1871)  &(1692) & (1835) &  (1802) & (1825)& (1911)\\ \hline
	Long-term &US  & UK & Germany  & France & Italy& India \\	
	interest rates &(1786)  & (1700) & (1788)  & (1746) &(1807)& (1722) \\	\hline
	Currencies&   US  & UK & Germany & France &  Austria  & Russia \\	
	in gold units &(1723) & (1723) & (1794)  & (1800)&(1813)  & (1814) \\	\hline
	Commodities & Oil & Silver & Copper & Wheat & Sugar & Cocoa \\	
	 &(1859) & (1718)&(1800)& (1808)   & (1784) & (1784) \\	\hline
\end{tabular}
\caption{List of financial markets used in the analysis of monthly data covering centuries.}	
\end{table}

\begin{figure}[t]\centering
	\includegraphics[height=5.5cm]{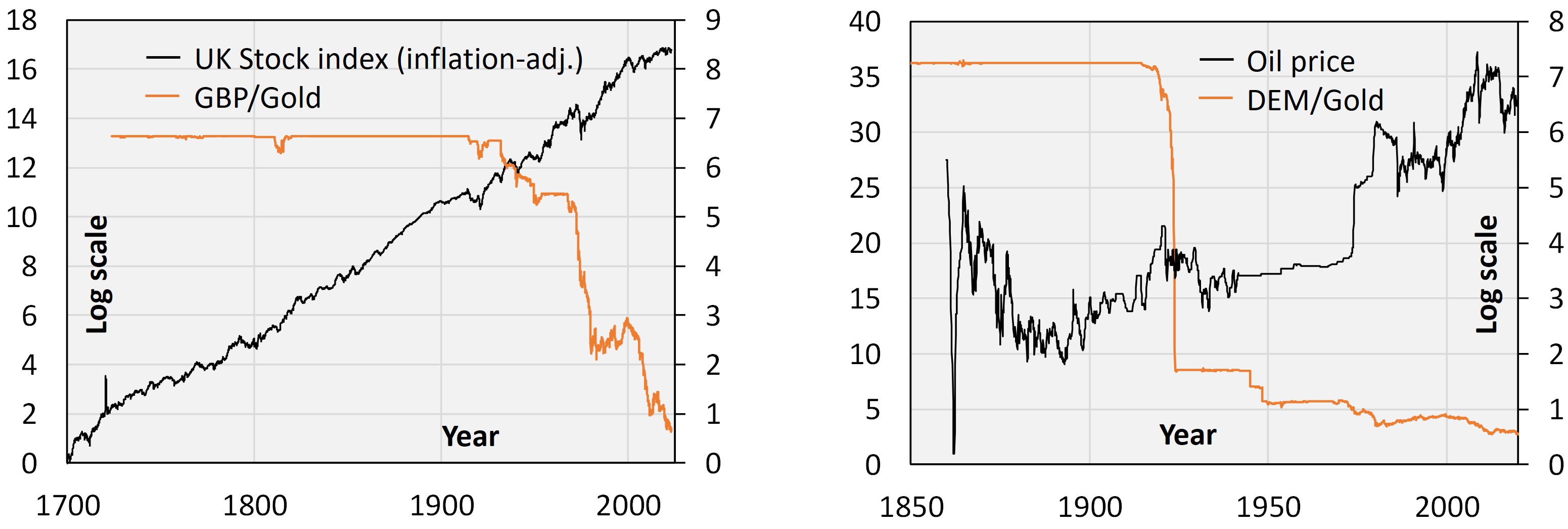} 
	\caption{Samples of our historical asset prices on a logarithmic scale to illustrate obstacles to quantitative analyses of historical data. 
Left: British stock market index and British pound in gold units since 1700. 
Right: Oil price and Deutsche Mark vs. gold since 1850.}
    \label{fig:monthly_data_issue}
\end{figure}

There are many issues with these data that have the potential to distort the results of our analysis. 
Let us illustrate them at a few examples; see also the review of the same data set in \cite{lemp}. 
\begin{itemize}
\item
Fig. 7 (left) shows the evolution of the British Pound in units of the gold price over 300 years on a logarithmic scale. The pound was pegged to gold over many decades. During these periods, the returns were zero and there were no trends, so no meaningful trend analysis is possible, except for very long-term trends. 
\item
Fig. 7 (right) shows the extreme downtrend of the Deutsche Mark in the 1920's due to German hyperinflation after World War 1. In fact, most currencies had periods of hyperinflation at some point in their history. Those start as normal down-trends but then culminate in "singularities", in the sense that the currencies devalue by orders of magnitudes within several months. These extreme events distort our analyses. We therefore discuss shorter-term currency trends  (up to 1 year) seperately below.
\item
Fig. 7 (left) also shows the inflation-adjusted evolution of the British stock market index since the late 17th century on a log scale. We observe a steady long-term equity risk premium of around 5\% over the centuries. Similar risk premia can be observed for all  other equity indices. These risk premia can be regarded as trends with infinite time horizon. We remove them before analyzing the remaining shorter-term trends.
\item
Apart from several equity market crashes, fig. 7 (left) shows a brief spike of the index around 1720, where its value briefly grew almost 10-fold, then crashed. This is not a data error. At that time the index consisted of only 3 companies, one of which - the South Sea Company - caused a huge speculative bubble that burst soonafter.  Such extreme outliers can also distort our analysis of shorter-term trends.
\item
Fig. 7 (right) also shows the evolution of the oil price since the industrial revolution on a log scale. Periods of huge price fluctuations and shocks are followed by periods of very rigid prices, e.g., in times when the oil price is politically controlled. This extreme range of volatility makes an overall analysis of oil price trends difficult. 
\item 
In fact, the volatilities of most of our assets have varied widely over the centuries. When normalizing returns to have standard deviation 1, we partly mitigate this problem by using an exponentially-weighted moving average volatility with a half-life of 30 years.
\item
There are also periods in history, during which prices were recorded imprecisely, e.g., as constant when they were only approximately constant. 
This makes the volatility seem zero, and thus leads to an infinite trend strength. We exclude such assets.
\item
For many other assets, the price was historically recorded only once a year or even only once in a decade (e.g., for rice in the 12th century in China, or for silk and safran in per-industrial times). 
We do not include such assets in our analysis either.
\end{itemize}

In view of these data issues with century-old data, in this section we will regard regression coefficients as statistically significant only if their t-statistics is $\ge3$.
We use trend horizons of $1.5\cdot 2^k$ months, with $k\in\{1,2,...,9\}$, corresponding to 3 months to 64 years.  
We use two data sets: one set of the monthly returns of the assets in table 8 since 1871, covering 150 years.
These data are more reliable than older historical data, and include all of the markets in table 8 during the whole time period,
except for the Nikkei index and for the shorter-term trends of currencies. We use them to analyze trends up to $k=7$ (16 years).
For $k=7,8$ (32, 64 years), we include each market in table 8 since the start of its track record. The longest track record, that of the British stock market index,
begins in 1692. These data are less reliable but contain just enough history to study trends that last a few decades.

\subsection{First Look at the Data}

\begin{figure}[t]\centering
	\includegraphics[scale=0.62]{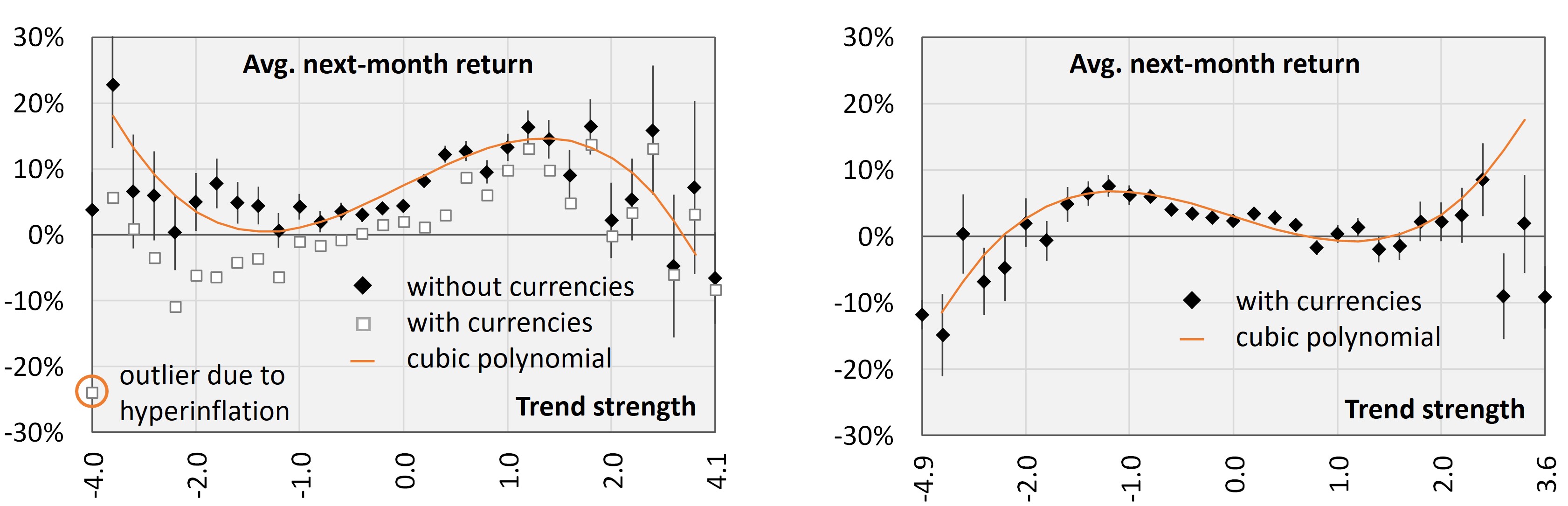}
	\caption{Average next-month return as a function of the current trend strength, based on 150 years of data. 
	Left: trend horizons up to 2 years. Right: horizons of more than 2 years.}
	\label{fig:minute_w}
\end{figure}

To explore what kind of pattern we can expect, 
we again start with a visualization of the data by grouping them into buckets of similar trend strengths, as for daily and minute data.
Here we use 5 buckets per unit of trend strength.
Fig. 8 (left) shows scatter plots of the average next-month return (on a log scale) as a function of a given month's trend strength for shorter-term trends.
We average over all years, all markets excluding currencies, and time horizons from 3 months to 1 year.\\

We clearly see the cubic polynomial that we have already observed for daily data since 1990 over these trend horizons. The positive linear and negative cubic coefficients
imply that weak trends tend to persist, while too strong trends tend to revert.
The new piece of information we gain is that this cubic pattern is not a recent phenomenon, but has existed for at least 150 years,
in line with previous results \cite{hurst,lemp,black}. \\

The white squares in fig. 8 (left) show the results if we include currencies, valued in units of the gold price. 
The whole curve is shifted downwards, because - over the decades - all currencies have lost value relatively to gold due to inflation.
The lower left data point (encircled) reflects the final months of periods of hyperinflation, as discussed above. \\

Fig. 8 (right) shows the same plot (including currencies) for long time horizons (4-16 years). The slope $b$ at the origin is clearly negative.
The cubic coefficient $c$ seems to be positive. We will confirm this by the regression analysis of the next sub-section. 
Thus, the scatter plot suggests that, for time horizons of many years, 
weak trends tend to revert. A possible interpretation is that this reflects economic cycles. On the other hand, strong long-term trends tend to persist. 
This may reflect the long-term rise and decline of individual economies, or of the supply and demand for key commodities such as oil or copper.  

\subsection{Regression Analysis}

We now carry out two cubic regression analyses with ansatz (\ref{quartic}) for monthly returns, 
one for shorter trend horizons of $T\in\{3,6,12\}$ months with all assets excluding currencies
(to avoid the hyperinflation outliers), and one for longer horizons $T\in\{2,4,8,16\}$ years including currencies. 
We begin with the period 1871-1923.

\begin{table}[h!]\centering
\begin{tabular}{ |p{2cm}||p{1.8cm}|p{1.8cm}|p{1.4cm} || p{1.8cm}|p{1.8cm}|p{1.4cm}| }	\hline
 	Parameter & \multicolumn{3}{|l||}{Trend horizons up to 1 year} & \multicolumn{3}{|l|}{Trend horizons $2-16$ years}\\ \hline
	{Coefficient}& Value  &Error &t-stat& Value  &Error &t-stat\\	\hline
	$a$   &+6.1\%    &$\pm 0.9$\% & 7.3  & $+2.2$\%     &$\pm 0.7$\% &  3.0\\  
	$b$ &$+7.0$\% & $\pm 1.4$\% &  $5.0$  & $-4.2$\%     &$\pm 0.9$\% & 4.8 \\	
	$c$ &  $-1.4$\%  &  $\pm 0.4$\% & $3.3$\  & $+1.1$\%     &$\pm 0.2$\% &5.2\\ \hline\hline
	{R-squared} & Single & \multicolumn{2}{|l||}{Aggregated}& Single & \multicolumn{2}{|l|}{Aggregated} \\ \hline
	$R^2$ &$11.0$\ bp & \multicolumn{2}{|l||}{15.4\ bp} &$5.0$\ bp & \multicolumn{2}{|l|}{8.5\ bp} \\	
	$R^2_{adj}$ &  $5.7$\ bp  &  \multicolumn{2}{|l||}{8.4\ bp} &  $2.2$\ bp  &  \multicolumn{2}{|l|}{4.4\ bp} \\ \hline
\end{tabular} 
\caption{Regression results for ansatz (\ref{quartic}), using long-term monthly (instead of daily) data.
Left: shorter horizons excluding currencies. Right: longer horizons including currencies.}	
\end{table}

The results are shown in table 8. They confirm and quantify our qualitative observations that the linear coefficient $b$ is positive for shorter horizons and negative for longer horizons, 
while the cubic coefficient $c$ is negative for shorter horizons and positive for longer horizons. 
These results are statistically significant; the errors are computed by bootstrapping with 500 samples.
The adjusted R-squared is computed by 12-fold cross validation. 
As reported in the appendix, a quadratic term would clearly be insignificant for shorter time horizons.
For longer horizons, its $t$-statistics is $t=2.4$, which is below our significance threshold of 3. 
In view of this, we do not add a quadratic term, although we cannot rule out its existence.\\

\begin{figure}[t]\centering
	\includegraphics[scale=0.63]{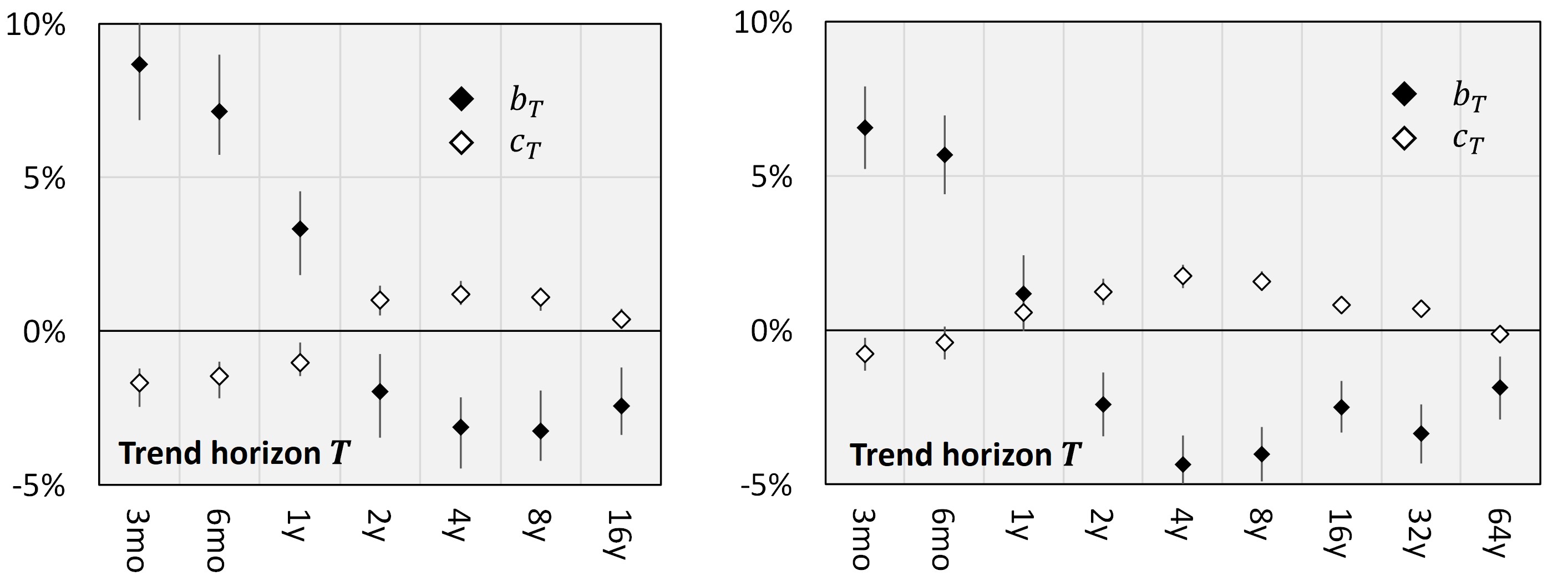} 
	\caption{Regression of monthly returns against trend strengths by time horizon, using ansatz (\ref{quartic}). Left: all assets since 1871, excluding currencies for $T\le1$ year.
	Right: extension of the analysis to 6 decades, based on each asset since the first year of its track record ($\ge1692$).}
	\label{fig:bc_century_1871}
\end{figure}

The persistence of trends $b$ and the strength of reversion $c$ are shown seperately for each time horizon in fig. 9 (left).
For horizons up to about two years, small trends indeed tend to persist ($b$ is significantly positive), 
while trends tend to revert when they become too strong ($c$ is negative). 
This re-confirms that 
this interplay of trends and reversion has characterized financial markets at least since the industrial revolution.
In fact, it may well go back to pre-industrial times, as shown in fig. 9 (right),
which is based on the longer (but less clean) set of monthly data since 1692.\\

At horizons of 1-2 years, the persistence of small trends disappears ($b\rightarrow0$), so the "trending regime" ends at these time scales.
For 2- and 4-year horizons, this has already been seen in our analysis of daily data since 1990. 
Our new observation for trend horizons of up to a few decades is that weak trends tend to revert (negative $b$), while
strong trends tend to persist (positive $c$).
As seen from fig. 9 (right), which extends to time horizons of up to 32 years,
this pattern has apparently characterized financial markets for centuries.
As noted above, it may reflect economic cycles, combined with 
long-term economic trends that last a decade or longer.\\

Our results are in close agreement with the 200-year-results \cite{black},
where it was also found that the parameters we call $b_T$ and $c_T$ both switch signs at time horizon $T\approx2$ years
(although we disregard the negative quadratic term).
Our results suggest that this reversion behavior extends from time scales of 5 years as in \cite{black} 
to at least a few decades, and that it has existed for at least 300 years.

\subsection{Refinements and Extension to Medieval Times}

To check if our reversion pattern for long trend horizons depends on the asset class, we  
have performed separate nonlinear regression analyses for equity indices, interest rates, commodities, and currencies. 
As we have already found in the case of daily data, for time horizons up to 1 year all asset classes show the same pattern: 
weak trends tend to persist, strong trends tend to revert.
Unfortunately, for longer-term horizons, the amount of noise in our 
monthly long-term data does not allow us to establish whether the pattern found above 
(reversion of weak trends, persistence of strong trends) occurs separately for each asset class.\\

However, on fundamental grounds, it is clear that there are differences between the asset classes. Equities differ by their long-term risk premia of around 5\% per annum. 
Interest rates have generally declined over the centuries and decades. 
Currencies have extreme down-trends in times of hyper-inflation,  
and commodity prices display periods of extreme volatility. 
Thus, the universality of the interplay of trends and reversion that we have observed from daily data since 1990 is unlikely to extend to time horizons of a few years and longer.\\

We have also tried to extend our observations on the interplay of trends and reversion all the way back to medieval times,
trying to extend linear results in \cite{grey}, but our data set is too limited to yield statistically significant {\it nonlinear} regression results.\\

\begin{table}[h!]\centering
\begin{tabular}{ |p{2.5cm}||p{3.5cm}|p{3.5cm}|p{3.5cm}|  }	\hline
 	Parameter & \multicolumn{3}{|l|}{Regression results ($a=$ intercept, $b=$ beta to trend strength)} \\ \hline
	{Coefficient}& Value  &Error &t-statistics\\	\hline
	$a$   & +1.0\%    &$\pm 2.0$\% & 0.6\\  
	$b$ &$-11$\% & $\pm 5.6$\% &  $2.0$ \\	 \hline
\end{tabular}
\caption{Linear regression results for yearly interest rates since the 14th century.}	
\end{table}

We could only do a simple linear analysis for interest rates, for which we have annual quotes for 
Germany (since 1327), France (since 1388), England (since 1311), and the Netherlands (since 1401).
There are also historical prices for gold and silver since 1258, but since gold and silver essentially {\it were} the currencies at the time,
these prices are not useful for a regression analysis: they tend to be stable over decades, 
interrupted by occasional jumps in times of inflation. 
Price quotes for more exotic assets such as silk or safran are too infrequent and too discontinuous
to be included in the analysis.
The linear regression for interest rates since the 14th century, with trend horizons 
from 2 to 128 years, is shown in table 9. Not surprisingly, it indicates that trends in interest rates that last several decades tend to revert.  

\section{Synthesis and Discussion} 

Let us now glue the observed values of $b$ and $c$ from intra-day data since 2010, daily data since 1990,
and long-term monthly data since 1692 together. 
Since these data cover different time periods, gluing the results together only makes sense and will only work if $b$ and $c$ are relatively stable across different time periods.\\

To make $b_T$ and $c_T$ for different time scales $T$ comparable, the time horizon $N$ of the future return $R_N$ should scale with the time horizon $T$ of the trend used to predict it.
We therefore rescale the coefficients $b_T,c_T$ to $\tilde b_T,\tilde c_T$ as in (\ref{redef}):
\begin{equation}
\tilde b_T=b_T\cdot\sqrt{T/60}\ ,\ \ \ \tilde c_T=c_T\cdot\sqrt{T/60}.
\end{equation}
This is equivalent to normalizing both the trend strength $\phi_T$ and the predicted returns $R_{T/60}$ to always have variance 1,
so they can be compared across all time scales from minutes to centuries. 
That is, we use the $T$-minute trend to predict the subsequent $T$-second return, 
the $T$-day trend to predict the subsequent $T/60$-day return, and the $T$-month trend to predict the subsequent $T/5$-year return.
This yields the consolidated scaling behavior of financial market trends shown in fig. \ref{fig:combined}.
In summary, we have observed the following:

\begin{figure}[t]\centering
	\includegraphics[height=5.2cm]{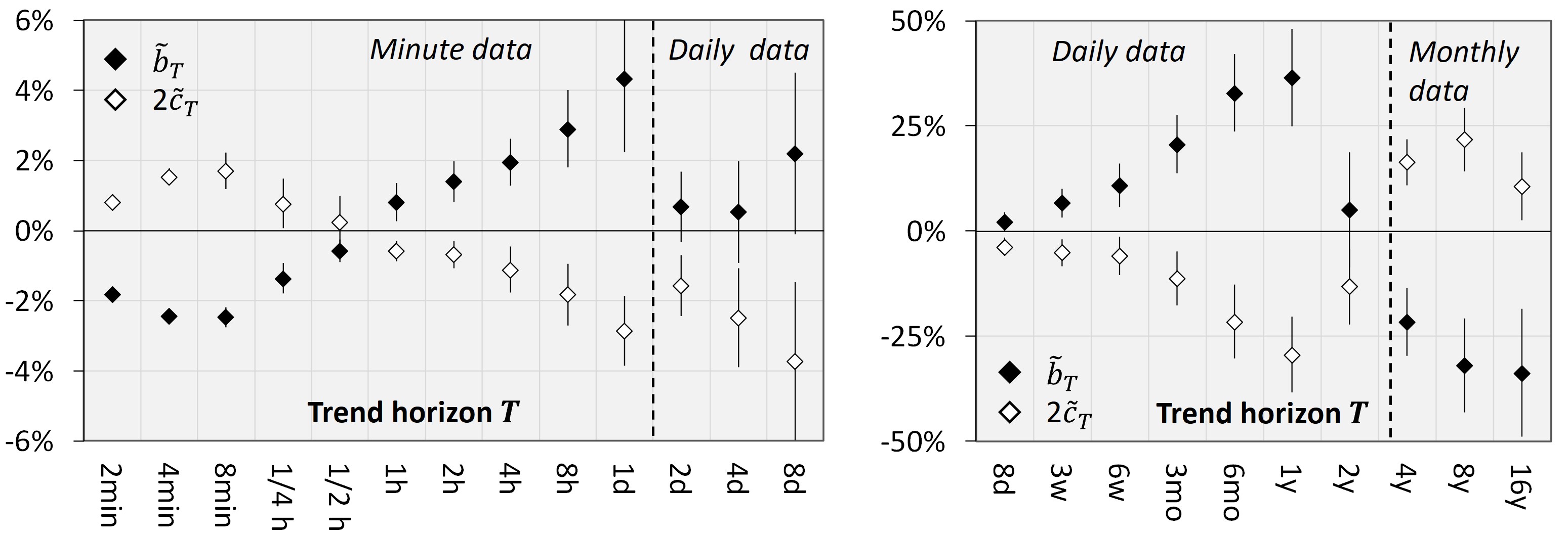} 
	\caption{Regression coefficients from minutes to decades based on ansatz (\ref{quartic}), combining (i) intraday data since 2010, (ii) daily data since 1990, and (iii) monthly data since 1692. 
Left: time horizons of $T\le8$ days. Right: time horizons $T\ge8$ days.
          Although the three data sets cover different time periods, the results connect with each other within error margins.}
	\label{fig:combined}
\end{figure}

\begin{itemize}
\item {\it Intraday time horizons:} up-ticks in futures prices (trade prices) tend to be followed by down-ticks, and vice versa.
This effect is stronger for shorter time horizons and can be explained as an artifact of the tick size. 
It may be used to reduce trading costs, but not as a profitable trading strategy (markets seem to be efficient at minute scales). 
\item {\it Time horizons up to 15 minutes:} even after subtracting this "tick size reversion", weak trends in financial markets generally tend to revert at these scales (negative $\tilde b$). 
This is not the case for stronger trends, which tend to persist on average (positive $\tilde c$), possibly due to fundamental causes of some trends.
In the regression analysis, quintic and possibly higher-order polynomials in the trend strength are statistically significant at these short time scales. 
Their coefficients are not universal - they depend on the asset.
\item {\it Time horizons from 30 minutes to a few hours:} after subtracting the tick size reversion effect, the sign of $\tilde b$ switches sign
and the pattern can be modeled well by a regression with universal linear and cubic terms ($\tilde b>0, \tilde c<0$), just as for daily data. 
At horizons of more than a few hours, $\tilde b$ is strong enough to overcome the tick size reversion effect. 
\item {\it Time horizons from a few hours to several days:} 
small trends tend to persist ($\tilde b>0$), while strong trends tend to revert ($\tilde c<0$).
In this range, $\tilde b$ and $\tilde c$ are small. The jump from minute- to daily data is still within estimation errors.
The data are consistent with assuming that $c$ is constant and financial markets are approximately scale invariant. 
$\tilde b$ and $\tilde c$ seem to be universal, i.e., independent of the asset or time zone (within errors).
This has been explained in \cite{me} by modeling trends as purely social phenomena that arise in the social network of investors,
which is assumed to be near a critical point.
\item {\it Time horizons from a few days to one year}: 
Trends in all markets still tend to persist as long as they are weak, and to revert when they become too strong.
However, $\vert \tilde b\vert,\vert \tilde c\vert$ grow with the time horizon and peak at 6-12 months. This pattern has long been exploited by trend-followers. It again seems to be 
universal (independent of the asset). 
\item {\it Time horizons from a few years to decades:} $\tilde b$ and $\tilde c$ become weaker and switch signs at horizons of $\approx2$ years, beyond which weak trends tend to revert
and strong trends seem to persist. This could reflect economic cycles and long-term developments.
\end{itemize}
These results provide a detailed set of empirical observations that measure the scaling behavior of financial markets.
Theoretical model of financial markets, such as \cite{bouchTV,lux, farmer, bouchaud, sornette, puk, puk1}, must replicate them, as well as other stylized facts \cite{mant,cont}. \\

Let us conclude by tentatively interpreting our results in the light of one such model, the lattice gas model \cite{me}.
The lattice represents the social network of investors, while the gas molecules represent the shares of an asset that are distributed across this network.
The order parameter $\phi$ corresponds to the deviation of the price of the asset from its long-term value.
Trend-followers and statistical arbitrageurs are expected to drive the lattice gas to its critical point, near which
it is described by scalar field theory with Landau potential
\begin{equation}
V={1\over2}\tilde b \phi^2-{1\over4}\tilde c\phi^4+{1\over6}\tilde d \phi^6+ ...,\ \ \ \text{where}\  \tilde b,\tilde d\ll1,\ \tilde c=-g^*.
\end{equation}
Here, it is assumed that "price-minus-value" $\phi$ can be measured as today's price minus a long-term average price,
which is essentially the (rescaled) trend strength $\phi_k$ in (\ref{return}) in the limit of large $T=2^k$, close to the correlation time. 
$g^*>0$ is the value of the $\phi^4$ coupling constant at its renormalization group fixed point, to which it tends (or, in renormalization group language, "flows") at large scales. 
The derivative $V'(\phi)$ plays the role of the cubic driving force in (\ref{quartic}).\\

$\tilde d$ is an {\it irrelevant} coupling constant in the renormalization group sense. This means that it shrinks towards zero at large scales $T>T_1$ for some $T_1$, but grows at small scales $T<T_1$, 
below which non-universal microscopic details of the molecules (corresponding to market details such as tick sizes and liquidity) become visible.
This matches our observations, if we identify $T_1\approx15-30$ minutes.\\

By contrast, $\tilde b$ is a {\it relevant} coupling constant in the renormalization group sense. This means that it shrinks to zero at {\it small} scales $T<T_2$ for some $T_2$, but grows at {\it large} scales $T>T_2$.
In the range $T_1<T<T_2$, the model is near the critical point. There, it should be approximately scale invariant and characterized by universal critical exponents, such as Hurst exponents.
Consistency with our observations requires a $T_2$ of roughly a few days.\\

$\tilde b<0$ implies that the lattice gas is slightly below the critical temperature. 
If the model is correct, $\tilde b$ should grow as a power $T^\beta$ of the time horizon $T$, until $T$ reaches the correlation time $\tau$, beyond which auto-correlations decay rapidly.
This is consistent with fig. 10 for $\tau\approx$ a few years. 
We can also estimate the critical exponent $\beta\approx 0.4-0.6$ from fig. 10. This value, as well as the Hurst exponents, must be replicated by the correct variant of the lattice gas model.
Further work on this potential interpretation of our results is underway.

\section*{Acknowledgements} 

We would like to thank Henriette (formely Wolfgang) Breymann, Ashkan Nikeghbali, Thomas Lehérici, and Maximilan S. Janisch for discussions, Gisela Reichmuth for
remarks on tick data, and Donat Maier for helping us to clean the intra-day and long-term data.
This research is supported by grant no. PT00P2\_206333 from the Swiss National Science Foundation. 

\section*{Appendix: Even Regression Terms}

For completeness, here we generalize the regression ansatz (\ref{quartic},\ref{quintic}) to also include the quadratic and higher-order polynomial terms:
\begin{equation}
R(t+1)=a+b\cdot  \phi(t)+c_2\cdot\phi^2(t)+c \cdot \phi^3(t)+c_4 \cdot \phi^4(t)+d \cdot \phi^5(t)+\epsilon(t+1).\label{general}
\end{equation}
The regression results for daily data are shown in table 10. The first 2 colums are the results of table 3 with only linear and cubic terms.
In the next 2 columns, we add the quadratic term. Its coefficient $c_2$ is not significant (t-statistics $<1$). 
It increases the in-sample $R^2$ (both for single and aggregated time scales), but it decreases the out-of-sample $R^2$. This is a classical sign of data overfitting.\\

The last 2 columns include all coefficients. The adjusted $R^2$ is similar as for the linear-plus-cubic regression.
However, the $t$-statistics for the coefficients $b-d$ are generally smaller, which can be attributed to overfitting outlier events with $\vert t\vert\ge2.5$. 
We conclude that the linear-plus-cubic model ist the most reliable one, 
although we cannot exclude that the response function may rise/fall more steeply than $\phi^3$ in the far tails.\vspace{3mm}

\begin{table}[h!]\centering
\begin{tabular}{ |p{1.5cm}||p{2.5cm}|p{1cm}|p{2.5cm}|p{1cm}|p{2.5cm}|p{1cm}| }	\hline
 	& \multicolumn{2}{|l|}{Linear \& cubic terms}&\multicolumn{2}{|l|}{With quadratic term}&\multicolumn{2}{|l|}{5$^{th}$ order polynomial} \\ \hline
	{Coeff.}& Value in \%  &t-stat & Value in \%&t-stat& Value in \%&t-stat\\	\hline
	$a$   & $+1.33\pm 0.41$ &  $3.3$&$+1.45\pm 0.36$ &  $4.0$&$+1.59\pm 0.30$ &  $5.3$\\  
	$b$ &$+1.29 \pm 0.43$ &  $3.0$ &$+1.30 \pm 0.43$ &  $3.0$ &$+0.33 \pm 0.44$ &  $0.7$ \\	
	$c_2$ &&&$-0.15 \pm 0.36$ &  $0.4$ &$-0.50 \pm 0.33$ &  $1.5$\\	
	$c$ &  $-0.62  \pm 0.23$ & $2.7$&$-0.63  \pm 0.23$ & $2.7$&$+0.36  \pm 0.36$ & $1.0$\\
	$c_4$ &&&&&$+0.07 \pm 0.12$ &  $0.6$\\	
	$d$ &&&&&$-0.15 \pm 0.07$ &  $2.1$ \\	 \hline\hline
	$R^2$ in bp & Single $T$ & Aggr.&Single $T$ & Aggr.&Single $T$ & Aggr. \\ \hline
	$R^2$ &$1.31$ & $4.91$&1.34&5.12&1.58&6.31  \\	
	$R^2_{adj}$ &  $1.03$  & $3.98$&0.84&3.50&0.96&4.04  \\ \hline
\end{tabular}
\caption{Extension of table 3 for daily returns to the general regression ansatz (\ref{general}).}	
\end{table}

Table 11 adds the quadratic term to the regression for the 150-year set of monthly data without currencies. 
For trends horizons up to one year, the quadratic coefficient $c_2$ is clearly insignificant, as for daily data.
For longer trends, it is negative with t-statistic 2.4, which is below our chosen significance threshold of $t=3$ for these long-term data. 
So we prefer not to add the quadratic regression term, although we cannot clearly rule it out either.

\begin{table}[h!]\centering
\begin{tabular}{ |p{1.5cm}||p{1.8cm}|p{1cm}|p{1.8cm}|p{1cm} || p{1.8cm}|p{1cm}|p{1.8cm}|p{1cm}|}	\hline
 	& \multicolumn{4}{|l||}{Trend horizons up to 1 year} & \multicolumn{4}{|l|}{Trend horizons $2-16$ years}\\ \hline
	{Coeff.}& Value (\%)  &t-stat&Value (\%)  &t-stat& Value (\%)  &t-stat&Value (\%)  &t-stat\\	\hline
	$a$ &$+6.1\pm 0.9$ & 7.3  &$+6.7\pm 0.9$ & 7.7  & $+2.2\pm 0.7$&  3.0&$+3.3\pm 0.7$&  4.6\\  
	$b$ &$+7.0\pm 1.4$ &  $5.0$ &$+7.0\pm 1.4$ &  $5.0$& $-4.2\pm 0.9$ & 4.8& $-3.9\pm 0.9$ & 4.3 \\	
	$c_2$ &&&$-0.2\pm0.7$& 0.3&&&$-1.1\pm0.5$ & 2.4 \\	
	$c$ &$-1.4\pm 0.4$& $3.3$&$-1.4\pm 0.4$& $3.3$& $+1.1\pm 0.2$&5.2&$+0.9\pm 0.3$&3.5\\ \hline\hline
	$R^2$ in bp & Single $T$ &Aggr.& Single $T$ &Aggr.& Single & Aggr.& Single $T$ &Aggr. \\ \hline
	$R^2$ &$11.0$ & 15.4 &$11.1$ & 15.6 &$5.0$& 8.5&7.5& 12.0 \\	
	$R^2_{adj}$ &  $5.7$  & 8.4&$3.5$  & 5.4&  $2.2$  & 4.4&3.3  &6.1 \\ \hline
\end{tabular} 
\caption{Extension of table 8 for monthly returns to the general regression ansatz (\ref{general}).}	
\end{table}

Table 12 shows an analogous analysis for intraday returns, in which case we also add a term $e\cdot\text{sign}(\phi)$ to ansatz (\ref{general}).
The quadratic/quartic terms are not or only barely statistically significant, and they do not increase the adjusted R-squared,
so we prefer to keep our regression model simple and drop these even terms as explanatory variables.

\begin{table}[h!]\centering
\begin{tabular}{ |p{1.5cm}||p{1.8cm}|p{1cm}|p{1.8cm}|p{1cm} || p{1.8cm}|p{1cm}|p{1.8cm}|p{1cm}|}	\hline
 	& \multicolumn{4}{|l||}{Trend horizons up to 16 minutes} & \multicolumn{4}{|l|}{Trend horizons 1 hour to 1 day}\\ \hline
	{Coeff.}& Value (\%)  &t-stat&Value (\%)  &t-stat& Value (\%)  &t-stat&Value (\%)  &t-stat\\	\hline
	$a$ &$+0.017$&  $0.8$     &0.055&2.8        & $+0.003$&  0.1            &$-0.008$&0.3\\   
	$b$ &$-0.912$& $13.6$      &-0.890&13.3        &$+0.132$ & 2.9             &$+0.159$&3.5\\
        $c_2$ &&   &-0.076&2.1       &&                &$-0.002$&0.1\\	
        $c$ &$+0.259$& $4.6$      &0.272&4.9        &$-0.039$ & 2.8               &$-0.038$&2.7  \\
        $c_4$ &&   &0.006&0.7       &&               &$-0.001$&0.2 \\
        $d$ &$-0.038$&$4.2$         &-0.042&4.6       &&                  &&\\	
        $e$ &$-0.282$&  $12.5$ &-0.287&12.5        &$-0.071$& 3.9&$-0.067$&3.7\\	 \hline\hline
	{$R^2$ in bp} & Single& Aggr. & Single& Aggr.              & Single& Aggr.                            & Single& Aggr. \\ \hline
	$R^2$ &$0.43$ & 0.56     &0.43&0.56        &0.012&0.025                    &0.013&0.027\\	
	$R^2_{adj}$ &  $0.42$   &0.55 &0.42&0.55&0.009&0.020          &0.007&0.016\\ \hline
\end{tabular}
\caption{Extension of tables 5,6 for intraday returns to the general regression ansatz (\ref{general}).}	
\end{table}


\begin{thebibliography}{9}\addtolength{\itemsep}{-2.5 pt} 

\bibitem{turtles} For a popular review, see, e.g., Covel, M., 2007. The Complete TurtleTrader: The Legend, the Lessons, the Results. Collins.

\bibitem{cutler} Cutler, D.M., Poterba, J.M. and Summers, L.H., 1991. Speculative dynamics. The Review of Economic Studies, 58(3), pp.529-546.

\bibitem{silber} Silber, W.L., 1994. Technical trading: when it works and when it doesn't. The Journal of Derivatives, 1(3), pp.39-44.

\bibitem{fung} Fung, W. and Hsieh, D.A., 1997. The risk in hedge fund strategies: Theory and evidence from trend followers. The review of financial studies 14, no. 2 (2001): 313-341.

\bibitem{jaeger} 
Jaeger, L. and Wagner, C., 2005. Factor modeling and benchmarking of hedge funds: can passive investments in hedge fund strategies deliver? The Journal of Alternative Investments, 8(3), pp.9-36. 

\bibitem{erb}
Erb, Claude B., and Campbell R. Harvey. "The tactical and strategic value of commodity futures." (2005).

\bibitem {miff} 
Miffre, J. and Rallis, G., 2007. Momentum strategies in commodity futures markets. Journal of Banking \& Finance, 31(6), pp.1863-1886.

\bibitem{shen}
Shen, Qian, Andrew C. Szakmary, and Subhash C. Sharma. "An examination of momentum strategies in commodity futures markets." Journal of Futures Markets: Futures, Options, and Other Derivative Products 27.3 (2007): 227-256.

\bibitem{mosk} 
Moskowitz, T.J., Ooi, Y.H. and Pedersen, L.H., 2012. Time series momentum. Journal of financial economics, 104(2), pp.228-250.

\bibitem{menk}
Menkhoff, Lukas, et al. "Currency momentum strategies." Journal of Financial Economics 106.3 (2012): 660-684.

\bibitem{baz}
Baz, J., Granger, N., Harvey, C.R., Le Roux, N. and Rattray, S., 2015. Dissecting investment strategies in the cross section and time series. Available at SSRN 2695101.

\bibitem{hurst}	
Hurst, B., Ooi, Y.H. and Pedersen, L.H., 2017. A century of evidence on trend-following investing. The Journal of Portfolio Management, 44(1), pp.15-29; 

\bibitem{lemp} 
Lempérière, Y., Deremble, C., Seager, P., Potters, M., and Bouchaud, J. -P. 2014. “Two Centuries of Trend Following.” Journal of Investment Strategies 3 (3): 41–61. 

\bibitem{grey}
Greyserman, Alex, and Kathryn Kaminski. Trend following with managed futures: The search for crisis alpha. John Wiley \& Sons, 2014.

\bibitem{black}
Bouchaud, Jean-Philippe, et al. "Black was right: Price is within a factor 2 of Value." Available at SSRN 3070850 (2017).

\bibitem{schmidhuber} 
Schmidhuber, Christof. "Trends, reversion, and critical phenomena in financial markets." Physica A: Statistical Mechanics and its Applications 566 (2021): 125642.

\bibitem{bouchTV}
Majewski, Adam A., Stefano Ciliberti, and Jean-Philippe Bouchaud. "Co-existence of trend and value in financial markets: Estimating an extended Chiarella model." Journal of Economic Dynamics and Control 112 (2020): 103791.

\bibitem{mant}
Mantegna, Rosario N., and H. Eugene Stanley. Introduction to econophysics: correlations and complexity in finance. Cambridge university press, 1999.

\bibitem{cont}
Cont, Rama. "Empirical properties of asset returns: stylized facts and statistical issues." Quantitative finance 1.2 (2001): 223.

\bibitem{me}
Schmidhuber, Christof. "Financial markets and the phase transition between water and steam." Physica A: Statistical Mechanics and its Applications 592 (2022): 126873

\bibitem{bak}
Bak, P., Tang, C. and Wiesenfeld, K., 1988. Self-organized criticality. Physical review A, 38(1), p.364.

\bibitem{me2}
Schmidhuber, Christof. "Critical Dynamics of Random Surfaces." arXiv preprint arXiv:2409.05547 (2024).

\bibitem{mendel}
Schmidhuber, Christof (2020), “Data for: Trends, Reversion, and Critical Phenomena in Financial Markets”, Mendeley Data, V1, doi: 10.17632/v73nzdt7rt.1

\bibitem{olson}
Müller, Ulrich A., et al. "Statistical study of foreign exchange rates, empirical evidence of a price change scaling law, and intraday analysis." J. of Banking \& Finance 14.6 (1990).

\bibitem{daco}
Gençay, R., Dacorogna, M., Muller, U. A., Pictet, O., \& Olsen, R. (2001). An introduction to high-frequency finance. Elsevier.

\bibitem{fama}
Fama, Eugene (1970). "Efficient Capital Markets: A Review of Theory and Empirical Work". Journal of Finance. 25 (2): 383. doi:10.2307/2325486. JSTOR 2325486. 

\bibitem{lux}
T. Lux, M. Marchesi, Scaling and criticality in a stochastic multi-agent model of a financial market, Nature 297 (1999) 498–500. 

\bibitem{farmer}
Farmer, J.D. and Joshi, S., 2002. The price dynamics of common trading strategies. Journal of Economic Behavior and Organization, 49(2), pp.149-171.

\bibitem{bouchaud}
Bouchaud, J.P. and Cont, R., 1998. A Langevin approach to stock market fluctuations and crashes. European Physical J. B-Condensed Matter and Complex Systems, 6(4). 

\bibitem{sornette}
Sornette, D., 2014. Physics and financial economics (1776–2014): puzzles, Ising and agent-based models. Reports on progress in physics, 77(6), p.062001.

\bibitem{puk}
Takayasu, M., Mizuno, T. and Takayasu, H., 2006. Potential force observed in market dynamics. Physica A: Statistical Mechanics and its Applications, 370(1), pp.91-97.

\bibitem{puk1}
Watanabe, K., Takayasu, H. and Takayasu, M., 2009. Random walker in temporally deforming higher-order potential forces observed in a financial crisis. Physical Review E, 80(5), p.056110.

\end{thebibliography}
\end{document}